\documentclass[10pt,twocolumn,letterpaper]{article}

\usepackage[pagenumbers]{cvpr} 

\definecolor{cvprblue}{rgb}{0.21,0.49,0.74}
\usepackage[pagebackref,breaklinks,colorlinks,citecolor=cvprblue]{hyperref}

\def\paperID{*****} 
\def\confName{CVPR}
\def\confYear{2024}

\usepackage{float}

\newcommand{\method}{\text{CT4D}} 

\title{CT4D: Consistent Text-to-4D Generation with Animatable Meshes}

\author{
    \normalsize Ce Chen$^{1}$ \quad Shaoli Huang$^{2\#}$ \quad Xuelin Chen$^{2}$ \quad Guangyi Chen$^{1,3}$ \quad Xiaoguang Han$^{4,5}$ \quad Kun Zhang$^{1,3}$ \quad Mingming Gong$^{1,6}$ \\
    \small $^{1}$Mohamed bin Zayed University of Artificial Intelligence \quad $^{2}$Tencent AI Lab \quad $^{3}$Carnegie Mellon University \quad \\
    \small $^{4}$FNii, CUHKSZ \quad $^{5}$SSE, CUHKSZ \quad $^{6}$University of Melbourne \\
}

\begin{document}
\twocolumn[{
    \renewcommand\twocolumn[1][]{#1}
    \maketitle
    \vspace*{-2.9em}
    \begin{center}
        \captionsetup{type=figure}
        \includegraphics[width=\linewidth]{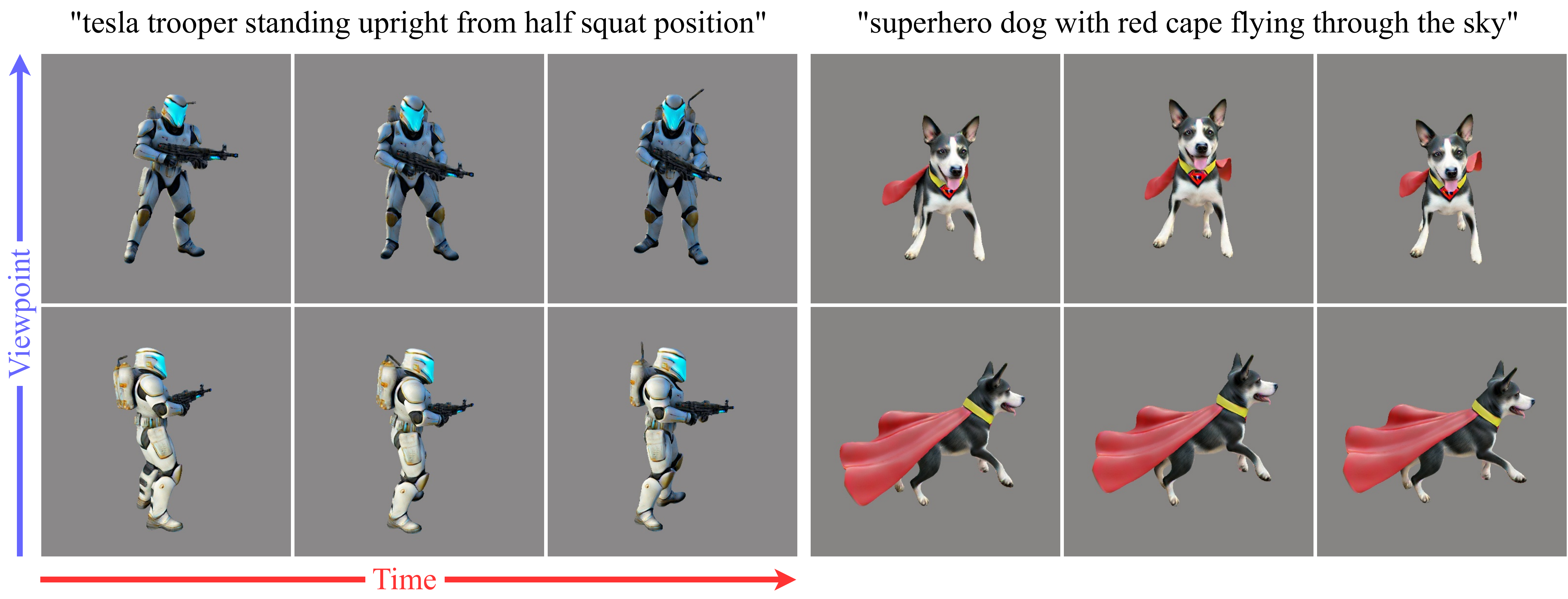} \vspace{-0.8em}
        \captionof{figure}{
            Generated samples from two text prompts viewed at different time steps and viewpoints. Video results are presented in the supplementary.
        }
        \label{fig:01}
    \end{center}
}]

\footnote{$\#$Corresponding authors.}

\begin{abstract}
Text-to-4D generation has recently been demonstrated viable by integrating a 2D image diffusion model with a video diffusion model.
However, existing models tend to produce results with inconsistent motions and geometric structures over time.
To this end, we present a novel framework, coined \method, which directly operates on animatable meshes for generating consistent 4D content from arbitrary user-supplied prompts.
The primary challenges of our mesh-based framework involve stably generating a mesh with details that align with the text prompt while directly driving it and maintaining surface continuity. Our \method{} framework incorporates a unique Generate-Refine-Animate (GRA) algorithm to enhance the creation of text-aligned meshes. To improve surface continuity, we divide a mesh into several smaller regions and implement a uniform driving function within each area. Additionally, we constrain the animating stage with a rigidity regulation to ensure cross-region continuity. 
Our experimental results, both qualitative and quantitative, demonstrate that our \method{} framework surpasses existing text-to-4D techniques in maintaining interframe consistency and preserving global geometry. Furthermore, we showcase that this enhanced representation inherently possesses the capability for combinational 4D generation and texture editing.
\end{abstract}

\section{Introduction}
The advent of diffusion models \cite{SohlDickstein2015DeepUL, Song2020ScoreBasedGM, nichol2021improved} across various modalities \cite{zhou2022magicvideo, singer2022make, ho2022imagen, rombach2022high, an2023latent, blattmann2023align, ge2023preserve, wang2023videofactory, wang2023lavie, shi2023mvdream, Qiu2023RichDreamerAG, Zt2vm, Wang2023SwapAI} has significantly advanced the development of text-to-4D generation. Utilizing score distillation sampling (SDS), recent research has demonstrated that diffusion models possess the necessary prior knowledge to generate 4D objects \cite{singer2022make, Singer2023TextTo4DDS, Bahmani20234DfyTG, Ling2023AlignYG, Zheng2023AUA}. Despite these advancements, the outputs of these methods frequently exhibit poor interframe consistency and fail to preserve geometry when adding motion.

\begin{figure}[htbp]
    \centering
    \includegraphics[width=0.75\linewidth]{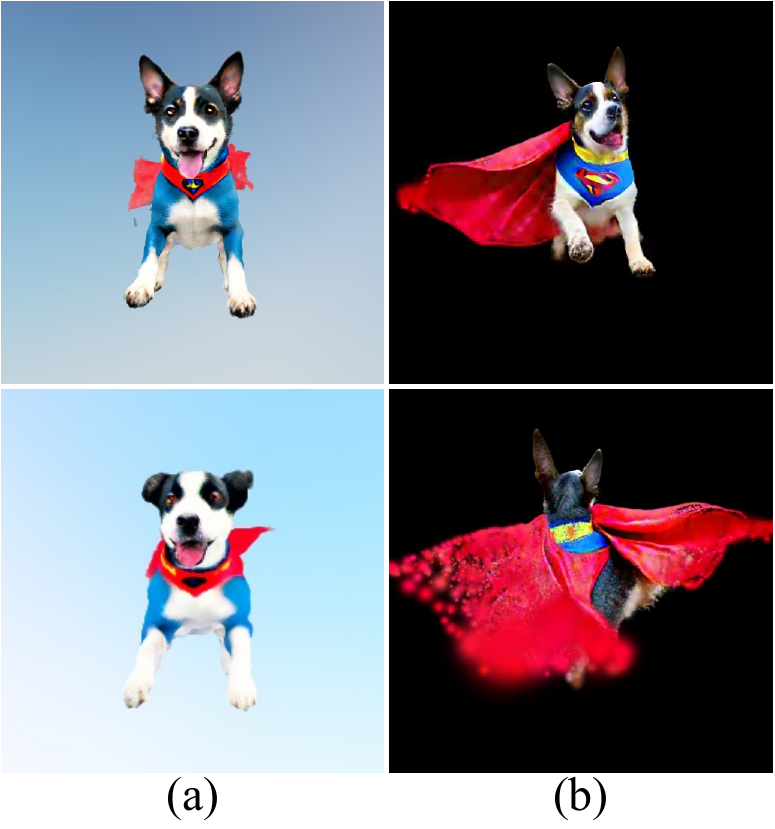}
    \caption{\textbf{Limitations of existing text-to-4D methods.} The text prompt is "superhero dog with red cape flying through the sky". (a) Front view outputs of 4D-fy with video SDS loss weights of 0.1 (1st row) and 1.0 (2nd row). The geometry and texture of the dog's ear degrade with larger weight due to not decouple the static and dynamic parts. (b) Front (1st row) and back (2nd row) view outputs of AYG. The front view looks fine, while the back view is distorted, with the cape appearing granulated.}
    \label{fig:02}
\end{figure}

Existing text-to-4D methods typically use dynamic neural radiance fields (NeRF) \cite{Mildenhall2020NeRF} as 4D representation \cite{singer2022make, Bahmani20234DfyTG, Zheng2023AUA}. However, the performance of these methods is frequently hindered by suboptimal geometry, which arises from the constraints inherent to text-to-video diffusion models. Thus, these dynamic approaches tend to exhibit inferior results compared to their static counterparts. Additionally, the failure to fully decouple static and dynamic components makes it difficult to preserve geometry at the dynamic stage, which leads to significant jitter in the edge regions of objects and difficulties in achieving interframe consistency. Consequently, the visual results rapidly deteriorate when incorporating more dynamic animations. Take 4D-fy \cite{Bahmani20234DfyTG} as an example, as illustrated in Figure \ref{fig:02} (a), with the larger video SDS loss weight, the geometry and texture of the dog degrade significantly, espacially around the ears. Furthermore, NeRF-based approaches demand substantial GPU memory, restricting generated video resolution and complicating subsequent editing.

Other text-to-4D methods \cite{Ling2023AlignYG} facilitate animations by adopting dynamic 3D Gaussian splatting (3DGS) \cite{Kerbl20233DGS}. While 3DGS-based methods successfully decouple static objects from dynamic parts, they often produce images with noticeable granularity due to the inherent particle nature of 3DGS. The generation results of align your gaussians (AYG) \cite{Ling2023AlignYG}, a 3DGS-based method, are granulated and distorted, like the cape shown in Figure \ref{fig:02} (b). Additionally, the 3D structure of these methods is often distorted, as 3DGS tends to overfit specific views. Furthermore, because 3DGS lacks practical constraints on geometry, objects can appear unnaturally deformed, leading to geometry preservation difficulty and low interframe consistency. Although 3DGS-based methods enable the quick combination of 4D objects, they do not disentangle geometry and texture, making texture modifications challenging. Please see Appendix \ref{sec:doem} for more discussions of existing methods.

To overcome these limitations, we propose a text-to-4D with animatable mesh (\method{}) framework, which generates animatable meshes with text-aligned details while preserving geometry, as shown in Figure \ref{fig:01}. Our unique Generate-Refine-Animate (GRA) algorithm makes this possible through a carefully designed three-stage process: (1) \textbf{Generating} a coarse NeRF using a text-to-multiview diffusion model \cite{shi2023mvdream} to ensure text alignment, then extracting a coarse mesh from the NeRF. (2) \textbf{Refining} the coarse mesh with multiple diffusion models \cite{rombach2022high, Qiu2023RichDreamerAG, shi2023mvdream} to produce geometry and texture with better details aligned with text. (3) \textbf{Animating} the refined mesh by converting it to an animatable form and optimizing it with a rigidity regulation \cite{SorkineHornung2007AsrigidaspossibleSM} for geometry preservation.

Unlike previous approaches \cite{singer2022make, Bahmani20234DfyTG, Ling2023AlignYG, Zheng2023AUA}, our \method{} framework employs animatable triangle mesh as the 4D representation, ensuring interframe consistency and mitigating jittering at the edges of the objects. Such explicit representation and decoupling between geometry and texture facilitate multiple object compositions and texture editing. To make a triangle mesh animatable, our \method{} framework first segments the mesh into multiple small regions and drives the mesh in a skeleton-free manner \cite{Liao2022SkeletonfreePT}, applying a uniform driving function for every vertex within each region. Then, when driving the mesh with text-to-video diffusion model \cite{Zt2vm, wang2023videofactory}, our \method{} framework constrains the animation with a rigidity regulation \cite{SorkineHornung2007AsrigidaspossibleSM} to maintain the surface continuity across regions.

We extensively evaluate our \method{} framework with various text prompts. The qualitative and quantitative results show that \method{} outperforms existing text-to-4D methods in interframe consistency and geometry preservation. Additionally, we show texture editing and multiple object composition results with the generated animatable meshes. In summary, \textbf{our contributions} include:
\begin{itemize}
    \item We propose a text-to-4D framework that leverages the animatable triangle mesh as the 4D representation. The approach effectively tackles the interframe consistency and geometry preservation limitations found in existing methods.
    \item We introduce the Generate-Refine-Animate (GRA) algorithm, which stably generates animatable meshes aligned with text prompts. This algorithm ensures the generation and animation of meshes with text-aligned details while maintaining surface continuity.
    \item Experimental results demonstrate that our \method{} framework outperforms current state-of-the-art methods (SOTAs) in terms of interframe consistency and geometry preservation.
\end{itemize}

\begin{figure*}[htbp]
  \centering
  \includegraphics[width=\linewidth]{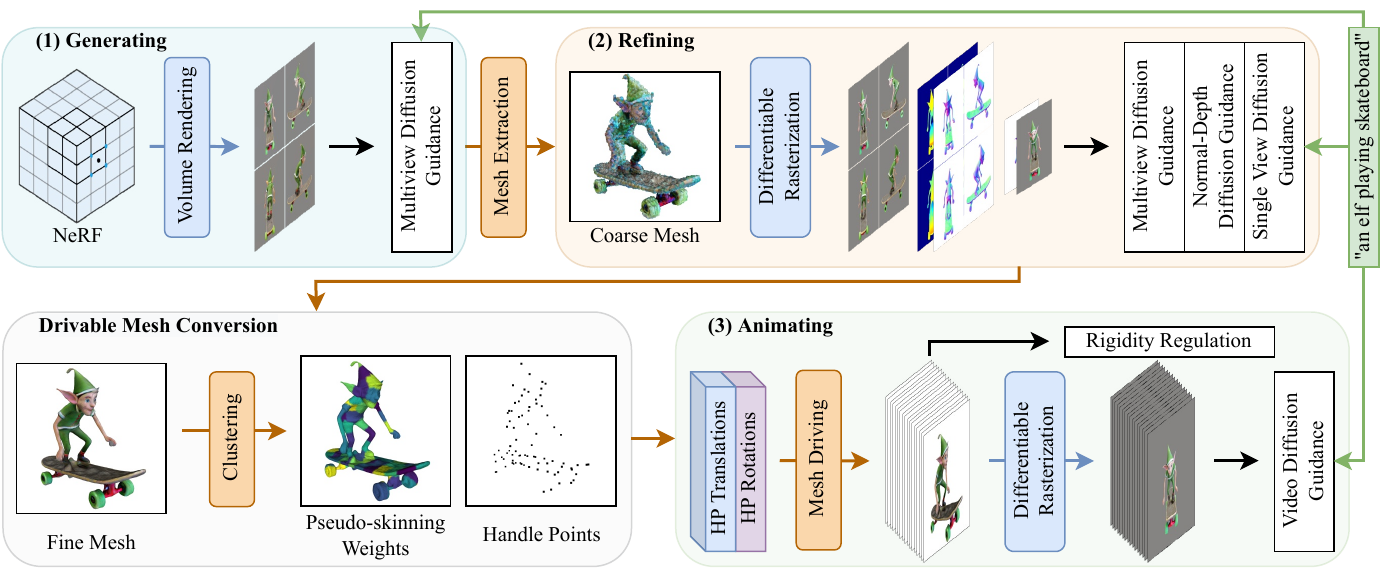}
  \caption{\textbf{\method{} Framework.} We generate animatable meshes using the Generate-Refine-Animate (GRA) algorithm. First, multiview images are rendered with a NeRF, and the SDS loss is calculated using a multiview diffusion model, resulting in a coarse static 3D object aligned with the text prompt. Next, a coarse mesh is extracted from the NeRF and sequentially optimized with multiple diffusion models (multiview, normal-depth, and single-view) to refine geometry and texture, producing a high-resolution mesh. Before animating, the refined static mesh is converted to an animatable form through clustering and generating pseudo-skinning weights and handle points. In the third stage, the mesh is driven by translations and rotations of the handle points (HP) and rendered into a sequence of keyframes to calculate the video SDS loss, thereby adding motion. Rigidity regulation is applied periodically to preserve the geometry during the animating stage.}
  \label{fig:03}
\end{figure*}

\section{Related Work}

\paragraph{Text-to-4D Generation.}
Recent methods have explored text-guided generation of dynamic 4D scenes using multiple diffusion models. Notable approaches include Make-A-Video3D (MAV3D) \cite{Singer2023TextTo4DDS}, 4D-fy \cite{Bahmani20234DfyTG}, and Dream-in-4D \cite{Zheng2023AUA}, which utilize dynamic NeRF \cite{Mildenhall2020NeRF} as their 4D representation. Meanwhile, Align Your Gaussians (AYG) \cite{Ling2023AlignYG} employs dynamic 3DGS \cite{Kerbl20233DGS} as its 4D representation. These dynamic NeRF and 3DGS methods are adept at modeling 4D scenes, enabling the generation of various dynamic objects with multiple diffusion models. However, MAV3D and 4D-fy are constrained by their video diffusion models due to the lack of decoupling between static and dynamic parts. AYG and Dream-in-4D have attempted to disentangle their 4D representations into a static 3D component and a deformation field. Still, they suffer from geometry preservation due to ineffective shape constraint methods. In contrast, our method leverages the animatable triangle mesh as the 4D representation, effectively decoupling geometry and texture. This explicit mesh attribute enables efficient shape constraint and the separation of static and dynamic parts.

\paragraph{Skeleton-free Mesh Deformation.}
Traditional skeletal rigging systems are the classical approach for driving static meshes \cite{MagnenatThalmann1989JointdependentLD}. However, skeletonization and rigging are complex processes that require expert intervention and cannot be automated easily. To address these limitations, skeleton-free mesh deformation methods have gained increasing attention. While these approaches typically rely on landmark annotations or rigging results \cite{Sumner2004DeformationTF, BenChen2009SpatialDT, Wang2019NeuralCF}, recent advancements have been made with methods like SfPT \cite{Liao2022SkeletonfreePT}, which generate consistent deformation parts across different character meshes and drives the mesh through these corresponding parts. Despite its advancements, SfPT requires the prediction of accurate skinning weights, which may not be available for arbitrary meshes. Our method overcomes this by using k-means clustering \cite{Hartigan1979AKC} to obtain pseudo-skinning weights. This simple yet effective approach enables us to drive a mesh without relying on any additional information. We achieve efficient mesh deformation by applying k-means clustering and SfPT to our generated triangle meshes, making the process more accessible for arbitrary meshes.

\section{Method}
\label{sec:m}

\begin{figure}[htbp]
    \centering
    \includegraphics[width=\linewidth]{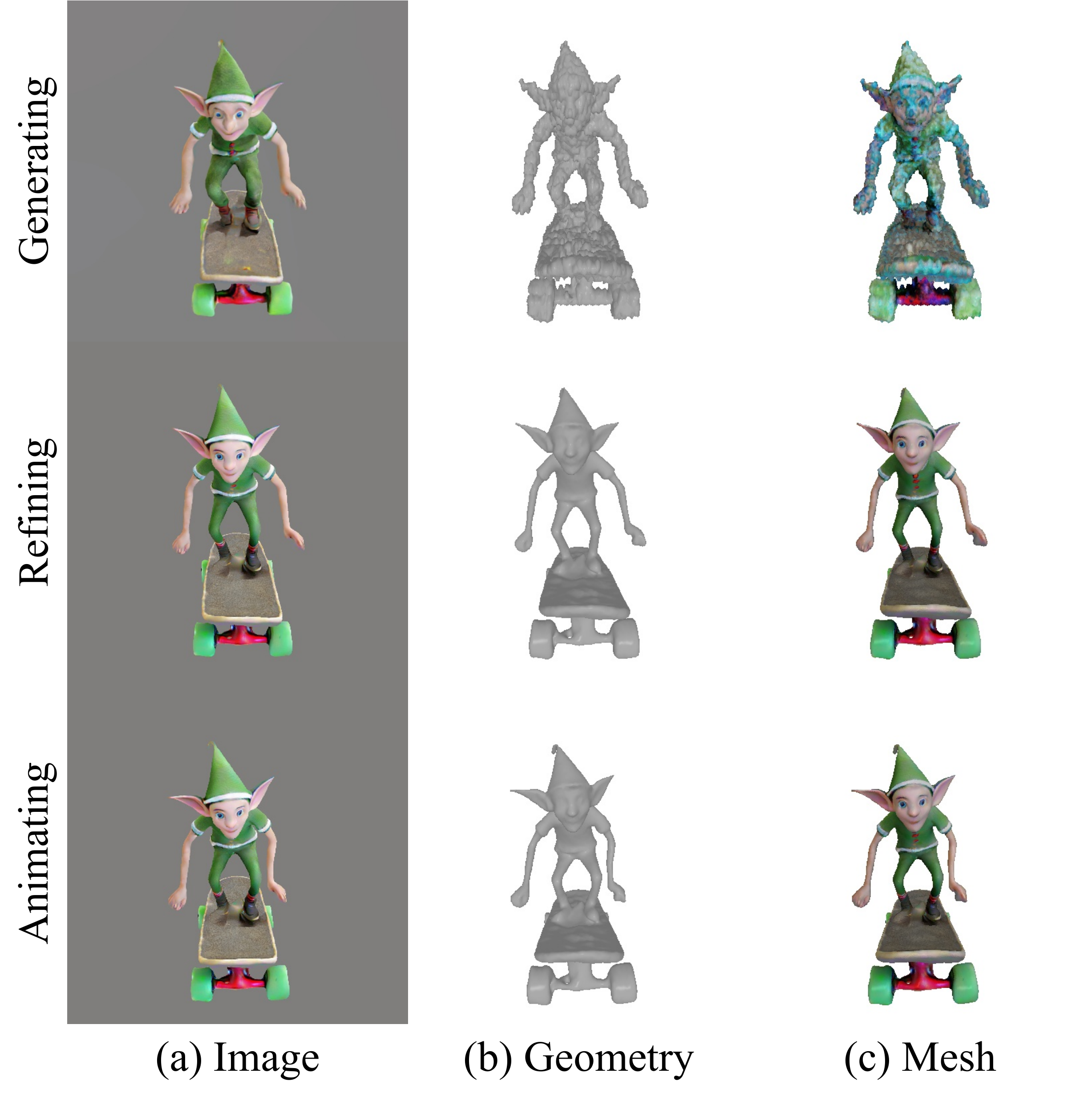}
    \caption{\textbf{Visual Results of GRA Algorithm.} The text prompt used is "an elf playing skateboard". (a) to (c) sequentially depict the image result, the geometry image rendered with the extracted mesh (without texture), and the final mesh image rendered with texture.
    }
    \label{fig:04}
\end{figure}

As illustrated in Figure \ref{fig:03}, given a text prompt, our \method{} framework generates an animatable mesh object aligned with the prompt. The key components of \method{} are the animatable mesh representation and the Generate-Refine-Animate (GRA) algorithm. Section \ref{sec:4r} introduces the animatable mesh as our 4D representation and explains how to convert an arbitrary triangle mesh into an animatable form simply yet effectively. In Section \ref{sec:gra}, we detail the GRA algorithm and describe the design motivation for each stage. We show how to use multiple diffusion models compositionally and adapt the GRA algorithm for mesh-based generation.

\subsection{4D Representation}
\label{sec:4r}
The 4D representation used in our \method{} framework is an animatable triangle mesh. We segment the mesh into multiple small regions using a clustering or segmentation method to drive an arbitrary mesh. Let $\phi$ be a mesh consisting of $N_v$ vertices denoted as $\mathbf{V}$ and $N_f$ faces denoted as $\mathbf{F}$. We use k-means clustering \cite{Hartigan1979AKC} due to its robustness and generality. K-means clustering is applied to the vertices $\mathbf{V}$ of the mesh $\phi$ to obtain $N_k$ clusters denoted as $\mathbf{K}$. Each vertex is associated with a $K$-dimensional pseudo-skinning weight corresponding to the $N_k$ clusters to drive the mesh. The pseudo-skinning weights $\mathbf{S} \in \mathbb{R}^{N_v \times N_k}$ are derived as follows:

\begin{equation}
s_{i,j}=\left\{
\begin{aligned}
1, \quad & \boldsymbol{v}_i \in \boldsymbol{k}_j \\
0, \quad & \boldsymbol{v}_i \notin \boldsymbol{k}_j \\
\end{aligned}, \quad \forall s_{i,j} \in \mathbf{S}
\right.,
\end{equation}

where $\boldsymbol{v}_i$ is the $i$-th vertex of mesh $\phi$ and $\boldsymbol{k}_j$ is the $j$-th cluster. The center of each cluster serves as a handle point for the mesh $\phi$, resulting in $N_k$ handle points denoted as $\mathbf{H}$ derived as:

\begin{equation}
\boldsymbol{h}_k=\frac{\sum_{i=1}^{N_v} s_{i, k}\boldsymbol{v}_i}{\sum_{i=1}^{N_v} s_{i, k}}, \quad \forall \boldsymbol{h}_k \in \mathbf{H}.
\end{equation}

The visualization of pseudo-skinning weights $\mathbf{S}$ and handle points $\mathbf{H}$ is shown in the animatable mesh conversion part of Figure \ref{fig:03}. Vertices within the same cluster are visualized with the same color. Each handle point governs the animation of the vertices within its associated cluster. Assuming there are $N$ keyframes in the animation, the motion can be expressed as a combination of translations $\boldsymbol{t} \in \mathbb{R}^{N \times N_k \times 3}$ and rotations $\boldsymbol{r} \in \mathbb{R}^{N \times N_k \times 4}$ for each handle point, with rotations represented using quaternions. Each frame's mesh $\phi_n$ is driven in a handle-based manner, denoted as $\mathcal{D}$. The vertices $\mathbf{V}_n$ and faces $\mathbf{F}_n$ of mesh $\phi_n$ are given by:

\begin{equation}
\begin{aligned}
\phi_n &= \mathcal{D}(\phi, n), \\
\boldsymbol{v}_{n,i} &= \sum_{k=1}^{N_k} s_{i, k}\left(\boldsymbol{R}_{n,k}\left(\boldsymbol{v}_i-\boldsymbol{h}_k\right)+\boldsymbol{t}_{n,k}+\boldsymbol{h}_k\right), \\
&\qquad \qquad \qquad \qquad \qquad \qquad \forall \boldsymbol{v}_{n,i} \in \mathbf{V}_n, \\
\boldsymbol{f}_{n,j} &= \boldsymbol{f}_j, \quad \forall \boldsymbol{f}_{n,j} \in \mathbf{F}_n,
\end{aligned}
\end{equation}

where $\boldsymbol{R}_{n,k}$ is the rotation matrix derived from $\boldsymbol{r}_{n,k}$, $\boldsymbol{v}_{n,i}$ is the $i$-th vertex of mesh $\phi_n$, $\boldsymbol{f}_{n,j}$ is the $j$-th face of mesh $\phi_n$, and $n$ represents the index of keyframes. Given a camera $c$, differentiable rendering \cite{laine2020modular}, denoted as $\mathcal{R}$ is used to render the animatable meshes $\Phi = \{\phi_1, \phi_2, \cdots, \phi_N\}$ into keyframes $I_{c} = \mathcal{R}(\Phi, c)$.

\subsection{Generate-Refine-Animate (GRA)}
\label{sec:gra}

Our \method{} framework aims to generate animatable mesh objects aligned with text prompts. To achieve this, we propose the Generate-Refine-Animate (GRA) algorithm, a compositional generation approach that progressively generates geometry, texture, and animations. The GRA algorithm is a core component of our \method{} framework. As illustrated in Figure \ref{fig:03}, our GRA algorithm consists of three key stages, contributing to text-aligned static 3D structure generation, high-quality mesh optimization, and mesh animation generation, respectively. In the following paragraphs, we detail each stage, explaining the motivation and corresponding function within the algorithm.

\begin{figure*}[htbp]
    \centering
    \includegraphics[width=\linewidth]{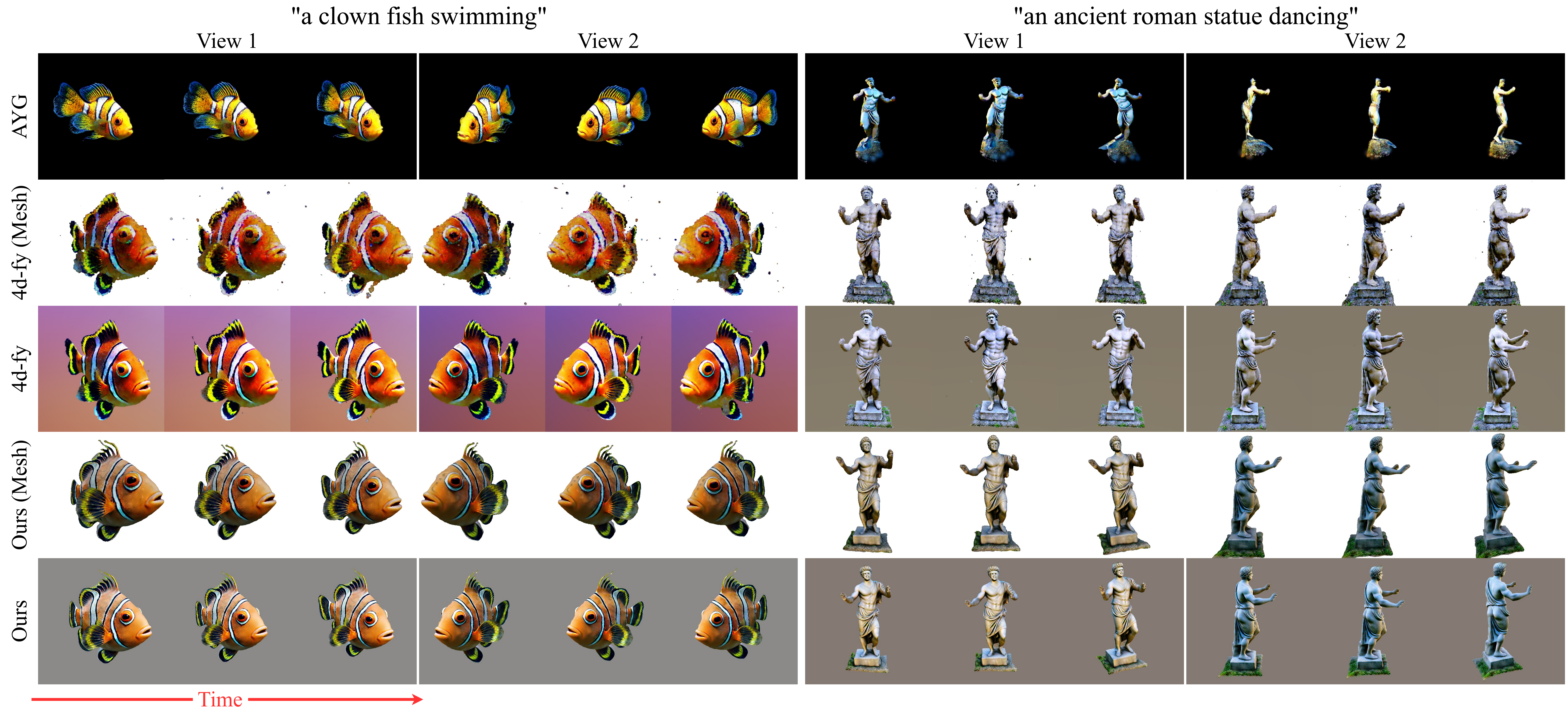}
    \caption{\textbf{Comparison with Existing Methods.} We compare our method with AYG and 4D-fy. To illustrate the geometry, we present the mesh rendering results from 4D-fy and our method in the 2nd and 4th rows, respectively. As the source code for AYG is not available, we cannot generate mesh rendering results for this method.}
    \label{fig:05}
\end{figure*}

\paragraph{Generating.}
Based on our observation of existing text-to-4D methods \cite{singer2022make, Ling2023AlignYG, Bahmani20234DfyTG}, NeRF is capable of generating text-aligned static 3D structure. Therefore, in this stage, we optimize a NeRF using multiview score distillation sampling (SDS) \cite{poole2022dreamfusion} with a diffusion model provided by MVDream \cite{shi2023mvdream} until convergence. For more details on SDS, please refer to Appendix \ref{sec:pre}. The number of iterations in this stage, denoted as $N_g$, is set to the same value as in \cite{shi2023mvdream}.

\paragraph{Refining.}
Text-aligned 3D structures from the generating stage often exhibit noisy geometry and texture, as seen in Figure \ref{fig:04}. To address this, we introduce a refining stage after generation, consisting of geometry and texture refinement. Results in Figure \ref{fig:04} demonstrate that refining surpasses the initial generation. In the geometry refinement phase, NeRF is further optimized using multiple diffusion models (multiview \cite{shi2023mvdream}, normal-depth \cite{Qiu2023RichDreamerAG}, and single-view \cite{rombach2022high}) via SDS for $N_{r_1}$ iterations. Subsequently, a tetrahedron mesh is extracted and optimized similarly for $N_{r_2}$ iterations. Afterward, the mesh is converted to triangles, with face number controlled for future optimization in animation. In the texture refinement phase, the triangle mesh is optimized with SDS of the multiview diffusion model and variational score distillation (VSD) \cite{wang2024prolificdreamer} of the single-view diffusion model for enhanced appearance. Geometry parameters are fixed during texture optimization. Refer to Appendix \ref{sec:pre} for details on VSD. The number of optimization iterations in this phase is denoted as $N_{r_3}$.

\paragraph{Animating.}
Before animating, we convert the triangle mesh obtained in the refining stage into an animatable form, as discussed in Section \ref{sec:4r}. In this stage, we initialize translation and rotation parameters for each handle point in every keyframe, representing rotations using quaternions. These parameters are optimized using a video diffusion model \cite{Zt2vm, Wang2023ModelScopeTT} through SDS for $N_{a_1}$ iterations.

To maintain surface continuity in cross-region areas, we integrate the as-rigid-as-possible (ARAP) method \cite{SorkineHornung2007AsrigidaspossibleSM} as a rigidity regulation step after every $N_{a_2}$ iterations of SDS. During rigidity regulation, the animation must be preserved while ensuring surface continuity. To achieve this, we sample $N_{fps}$ vertices using farthest point sampling (FPS), denoted as $\mathbf{V}_{fps}$. The loss for rigidity regulation is calculated as follows:

\begin{equation}
\begin{aligned}
    L_{rig} =& \lambda_{1} \Sigma_{n=1}^{N}{E(\phi_0, \phi_n)} \\
    &+ \lambda_{2} \Sigma_{n=1}^{N}{\text{MSE}(\mathbf{V}_{fps_n}^{\prime}-\mathbf{V}_{fps_n})},
\end{aligned}
\end{equation}

where $\phi_n$ is the mesh of $n$-th key frame, $\phi_0$ is the static mesh obtained from refining stage, $\mathbf{V}_{fps_n}$ represents the positions of $\mathbf{V}_{fps}$ in mesh $\phi_n$, $\mathbf{V}_{fps_n}^{\prime}$ represents the regulated positions of $\mathbf{V}_{fps_n}$, $\lambda_{1}$ and $\lambda_{2}$ are the weights for each term. The first term in $L_{rig}$ is the ARAP energy function, which regulates surface continuity. Please refer to Appendix \ref{sec:pre} for more details on this energy function. The second term maintains animation fidelity. Rigidity regulation is performed for up to $N_{a_3}$ iterations until convergence each time.

\section{Experiments}
\label{sec:e}

\subsection{Evaluation Settings}
\label{sec:es}

\paragraph{Metrics.}

\begin{figure*}[htbp]
    \centering
    \includegraphics[width=\linewidth]{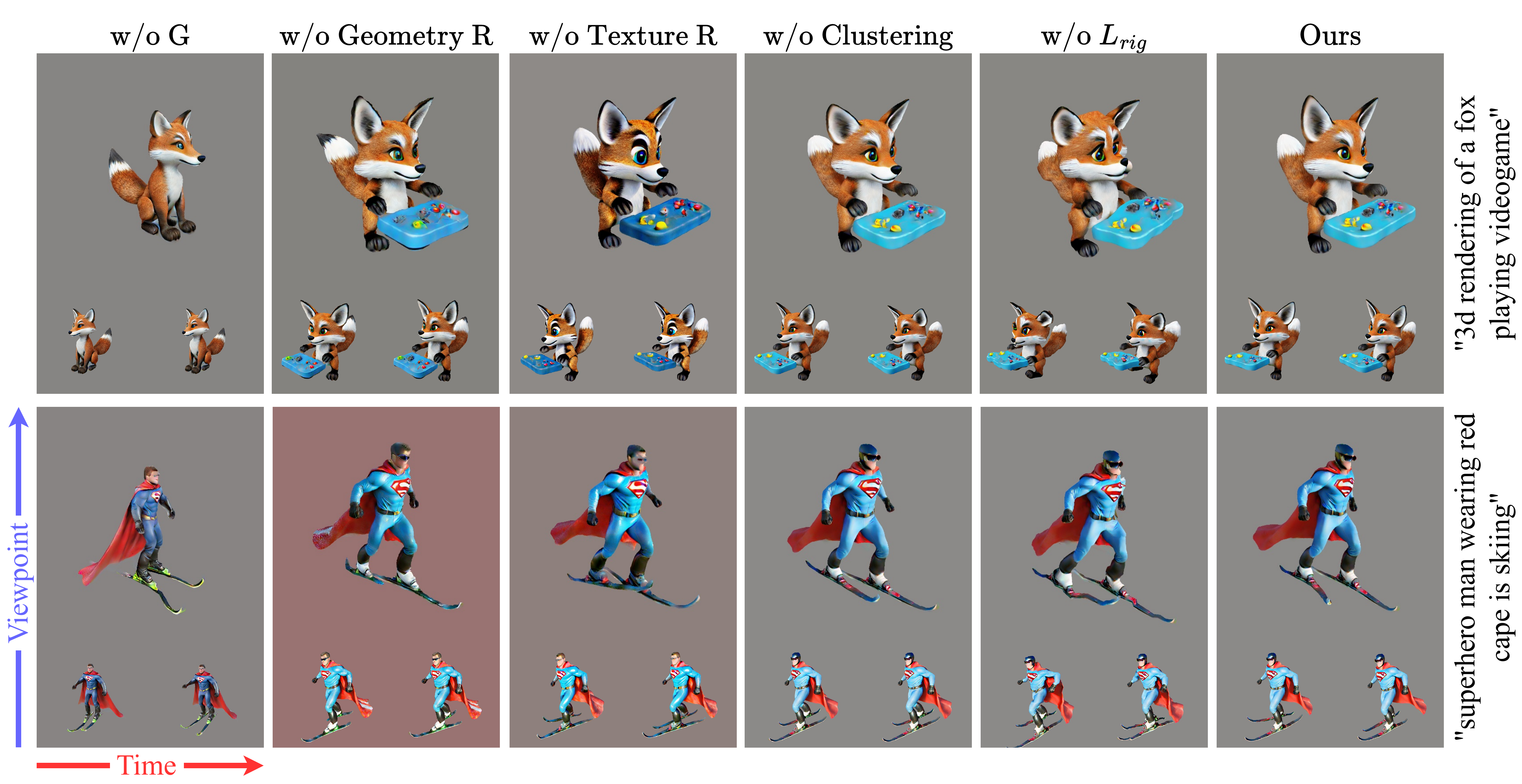}
    \caption{\textbf{Ablation Study.} We evaluate the impact of each component in our method by removing them from our framework. Without the generating stage (w/o G), the results do not align well with the text prompts. Without geometry refining (w/o Geometry R), the 3D structure is compromised. The appearance lacks realism without texture refining (w/o Texture R). The animation is unnatural without clustering for the mesh (w/o Clustering). The surface continuity is disrupted without rigidity regulation (w/o $L_{rig}$).}
    \label{fig:06}
\end{figure*}

We use the CLIP score \cite{liu2023one} and conduct a user study as our evaluation metrics. To assess interframe consistency (IC), we adopt the approach from \cite{Shi2024MotionI2VCA}, computing the cosine similarity between consecutive frames in the CLIP embedding space. Additionally, we calculate optical flow between the first frame and subsequent frames, determining the mean motion magnitude as average displacement (Disp.) to show the level of movement. Since the background in text-to-4D content is usually static, we calculate the average displacement both with and without the background (BG) by masking pixels with zero magnitude or not. Beyond evaluating RGB videos, we also calculate the CLIP score for videos of depth maps (Depth), normal maps (Normal), and rendered images of exported meshes without (Geometry) and with (Mesh) texture to assess geometry preservation. Each frame of the video is scored using CLIP ViT-B/32.

Additionally, we designed a user study for evaluation following 4D-fy. Randomly selected textual descriptions are presented to evaluators along with the results of our method and comparison methods. Evaluators are asked to express their preferences based on specific attributes of the videos. The original attributes in 4D-fy include Appearance Quality (AQ), 3D Structure Quality (SQ), Motion Quality (MQ), and Text Alignment (TA). To evaluate the interframe consistency and geometry preservation, we have included four additional attributes: Interframe Consistency (IC), Depth Map Quality (DMQ), Normal Map Quality (NMQ), and Extracted Mesh Quality (EMQ). Please refer to Appendix \ref{sec:usd} for more details on the user study.

\paragraph{Baselines and Prompts.} AYG and 4D-fy are the most latest methods for text-to-4D generation. Thus, we compare our \method{} framework with them. However, since AYG does not provide codes for reproducing the results in the paper, we selected 13 prompts from AYG for comparison. Additionally, we selected another 9 additional prompts for conducting ablation experiments and comparing different components of our method, following 4D-fy.

\paragraph{Implementation Details.} In our framework, the threestudio framework \cite{threestudio2023} is utilized to ensure reproducibility. The number of clusters $N_k$ is set to $80$. The number of keyframes $N$ is set to $16$. The value of $N_g$, $N_{r_1}$, $N_{r_2}$, $N_{r_3}$, $N_{a_1}$, $N_{a_2}$, and $N_{a_3}$ are set to 10000, 2000, 2000, 25000, 30000, 500, and 500, respectively. The sampling number of FPS $N_{fps}$ is set to $0.1 \cdot N_{v}$, where $N_{v}$ is the vertices number of mesh. $\lambda_{1}$ and $\lambda_{2}$ in the loss for rigidity regulation are $0.0001$ and $1.0$. We optimized the model on an NVIDIA A100 80G GPU. The approximate optimization time for each stage of GRA is 2, 3.5, and 10 hours, respectively. Please refer to Appendix \ref{sec:id} for more details.

\subsection{Comparison with Existing Methods.}
\label{sec:cwem}

\begin{table*}[htbp]
    \centering
    \resizebox{\linewidth}{!}{
    \begin{tabular}{l|cc|cccccc|ccccccccc}
        \toprule
        & \multicolumn{2}{|c|}{\textbf{Average Disp.}} & \multicolumn{6}{|c|}{\textbf{CLIP Score}} & \multicolumn{9}{c}{\textbf{User Study}} \\
        \textit{Method} & w/o BG & w/ BG & IC & RGB & Depth & Normal & Geometry & Mesh & AQ & SQ & MQ & TA & IC & DMQ & NMQ & EMQ & Overall \\ \midrule
        AYG & \textbf{7.15} & 3.23 & \underline{99.33} & 31.92 & \text{-} & \text{-} & \text{-} & \text{-} & 0.133 & 0.148 & \underline{0.257} & \underline{0.234} & 0.163 & \text{-} & \text{-} & \text{-} & 0.183 \\
        4D-fy & 4.16 & \textbf{3.84} & 99.20 & \textbf{32.87} & 24.80 & 26.17 & 24.40 & 30.80 &\underline{0.254} & \underline{0.237} & 0.195 & 0.210 & \underline{0.219} & 0.136 & 0.104 & 0.118 & \underline{0.186} \\
        \textbf{Ours} & \underline{6.09} & \underline{3.50} & \textbf{99.41} & \underline{32.79} & \textbf{26.48} & \textbf{28.76} & \textbf{27.68} & \textbf{31.74} & \textbf{0.612} & \textbf{0.615} & \textbf{0.547} & \textbf{0.556} & \textbf{0.618} & \textbf{0.864} & \textbf{0.896} & \textbf{0.882} & \textbf{0.630} \\ \midrule
        \textit{Ablation Study} & \multicolumn{17}{l}{~} \\ \midrule
        w/o G & 5.94 & 2.54 & 99.20 & 30.66 & 28.42 & 29.73 & 28.62 & 30.34 & 0.141 & \underline{0.162} & \underline{0.154} & 0.128 & 0.137 & 0.115 & 0.115 & 0.137 & 0.124 \\
        w/o Geometry R & 6.65 & \underline{3.46} & 99.21 & 30.65 & 27.93 & 28.46 & 27.85 & 30.53 & 0.081 & 0.081 & 0.115 & 0.090 & 0.090 & 0.115 & 0.111 & 0.077 & 0.085 \\
        w/o Texture R & 6.72 & 3.36 & 99.28 & 30.64 & \underline{29.10} & \textbf{29.82} & \underline{29.10} & 30.50 & \underline{0.167} & 0.141 & 0.145 & \underline{0.150} & 0.145 & \underline{0.154} & 0.150 & 0.154 & \underline{0.154} \\
        w/o Clustering & 6.40 & 3.18 & \textbf{99.48} & \underline{30.86} & 28.97 & 29.73 & 29.09 & \underline{30.78} & 0.111 & 0.150 & 0.098 & 0.120 & \underline{0.154} & \underline{0.154} & \underline{0.158} & \underline{0.158} & 0.145 \\
        w/o $L_{rig}$ & \textbf{7.14} & \textbf{3.67} & 99.05 & 30.83 & 28.09 & 28.85 & 27.96 & 30.67 & 0.051 & 0.060 & 0.081 & 0.094 & 0.064 & 0.085 & 0.081 & 0.081 & 0.068 \\
        \textbf{Ours} & \underline{7.00} & 3.44 & \underline{99.34} & \textbf{30.94} & \textbf{29.17} & \underline{29.80} & \textbf{29.11} & \textbf{30.84} & \textbf{0.449} & \textbf{0.406} & \textbf{0.406} & \textbf{0.419} & \textbf{0.410} & \textbf{0.376} & \textbf{0.385} & \textbf{0.393} & \textbf{0.423} \\ \bottomrule
    \end{tabular}
    }
    \caption{\textbf{Quantitative Results.} We evaluate the comparison with AYG and 4D-fy methods, as well as conduct an ablation study, using CLIP score and a user study. In all metrics, larger values indicate better performance. For the user study, the attribute values are normalized to sum up to 1. Since AYG only provides RGB video, certain metrics are not applicable to this method. The \textbf{best}-performing method is indicated in bold for each metric, while the \underline{second-best} method is underlined if there are more than three methods being compared.}
    \label{tab:01}
\end{table*}

When comparing the overall generation quality, we evaluate the final results of the same prompt output using different methods. As shown in Figure \ref{fig:05}, 4D-fy exhibits noticeable jitter at the edges between frames, and the exported meshes are noisy. AYG produces grainy images, some of which exhibit significant and unreasonable distortions. In contrast, our method generates results with smoother object edges, improved interframe consistency, and better mesh quality. Please see Appendix \ref{sec:aqr} for more qualitative results.

The CLIP score and user study results in Table \ref{tab:01} demonstrate that our method outperforms the other two methods in terms of interframe consistency and geometry preservation. Although the CLIP score of the RGB video from our method is slightly lower than that of 4D-fy, the CLIP scores for videos of depth maps, normal maps, and exported meshes from our method are all significantly higher, indicating superior geometry preservation. In terms of interframe consistency, our method achieves the highest score, and the average displacement of our method is comparable to the other methods, both with and without the background. We attribute the superior performance of our method to its design, which is specifically centered around the mesh. The inherent surface modeling attribute of the mesh enables us to generate 4D results with better interframe consistency.

\subsection{Ablation Study}
\label{sec:as}
To assess the importance of each component in our \method{} framework, we conducted ablation study by removing individual components. The quantitative and qualitative results of these experiments are presented in Table \ref{tab:01} and Figure \ref{fig:06}.

\paragraph{Generating Stage (w/o G).}
While generating a static 3D object without the generating stage is theoretically feasible, the resulting 3D objects often do not align well with the text prompts. Skipping the generating stage can lead to missing crucial parts of the object. For example, in the first case shown in Figure \ref{fig:06}, the video game controller is missing.

\paragraph{Refining Stage (w/o Geometry R and w/o Texture R).}
Although the generating stage produces text-aligned results, the geometry and texture are often coarse. Without geometry refinement, some parts may be missing, affecting realism. For example, in the second case, the ski board is broken. Similarly, without texture refinement, the appearance lacks realism, such as multiple eyebrows on the fox's face. The CLIP score of the normal map video without geometry refinement is the lowest, while without texture refinement, the CLIP score of the RGB video is the lowest.

\paragraph{Mesh Clustering (w/o Clustering).}
To assess the effectiveness of k-means clustering to convert the mesh into an animatable form, we conducted an experiment where we removed the clustering step and directly predicted the motion for all mesh vertices. The results, shown in Figure \ref{fig:06}, demonstrate that without clustering, the generated meshes exhibit unnatural deformation. This outcome underscores the importance of clustering in driving the mesh to achieve the motion described in the text prompt. Without clustering, the motion is minimal, resulting in frames that are nearly identical. Consequently, interframe consistency is the highest, while the average displacement is the smallest.

\paragraph{Rigidiy Regulation (w/o $L_{rig}$).}
To evaluate the significance of rigidity regulation, we performed the animating stage without $L_{rig}$. The experimental results in Figure \ref{fig:06} clearly illustrate the impact of this omission. Without $L_{rig}$, surface continuity cannot be maintained at the junctions of each region, resulting in a lack of coherence in the overall motion of the object. Without $L_{rig}$, the average displacement is the largest, while the interframe consistency is the lowest.

\begin{figure*}[htbp]
    \centering
    \includegraphics[width=0.9\linewidth]{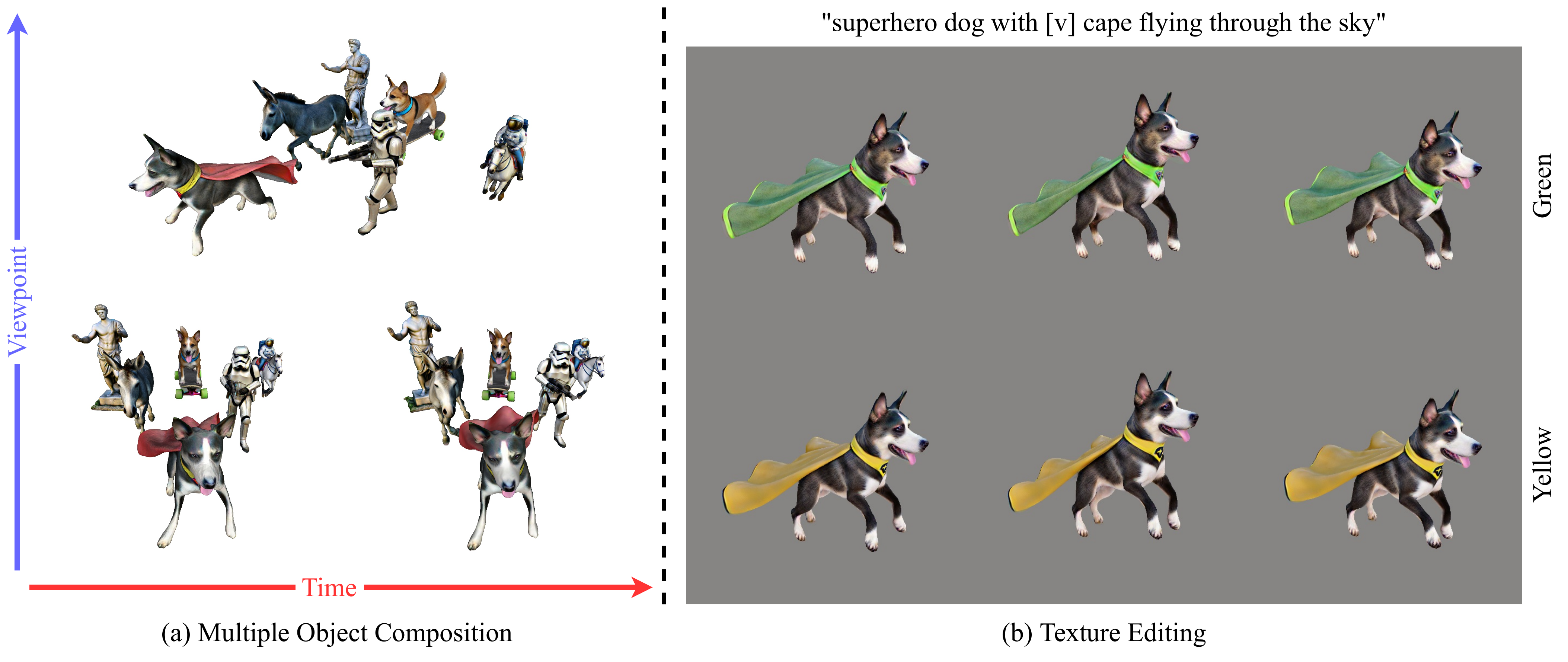}
    \caption{\textbf{Applications.} We show the results of multiple object composition (left) and texture editing (right) using animatable meshes generated with our \method{} framework.}
    \label{fig:07}
\end{figure*}

\subsection{Applications}
\label{sec:a}
The explicit property of the animatable mesh in our framework offers several advantages, including the ability to combine multiple dynamic objects and perform texture editing. These capabilities are not possible with the implicit NeRF-based approach. Although the 3DGS-based method allows for the combination of multiple dynamic objects, it fails to decouple the appearance and geometry of the objects, making texture editing challenging. Our \method{} framework overcomes this limitation, providing a more user-friendly solution for subsequent editing tasks. The explicit mesh representation enables efficient and flexible editing of object surfaces, enhancing the overall usability and versatility of our method. As shown in Figure \ref{fig:07}, the results of multiple object composition are obtained by arbitrarily placing many animatable meshes generated with our method. For texture editing, since our method disentangles texture and geometry and decouples static and dynamic parts, we only need to rerun the texture refining with the new text prompts without changing the geometry and animation. Please refer to Appendix \ref{sec:aqr} for more results.

\section{Conclusion}
\label{sec:c}

We propose \method{}, a framework for text-to-4D generation based on animatable mesh representation.
By leveraging the animatable mesh as our 4D representation, we enable the direct generation of animatable mesh objects while maintaining interframe consistency. To generate animatable meshes with text-aligned geometry and texture, our \method{} framework employs the Generate-Refine-Animate (GRA) algorithm to enhance the generation process. Additionally, mesh vertex clustering and rigidity regulation ensure surface continuity during the animating stage. Our experimental results demonstrate that impressive generation outcomes can be achieved using existing open-source models with our framework. Furthermore, we showcase the applications enabled by the explicit mesh representation, offering valuable solutions for subsequent content creation tasks. We will release the complete code for our \method{} framework upon paper acceptance to facilitate further advancements in this field. Our work advances text-to-4D mesh object generation and opens new possibilities for creative applications across various domains.

\paragraph{Limitations}
Our method has several limitations that should be addressed in future work. Firstly, when generating the static mesh, we do not account for joint movements, which are crucial for animation. This oversight limits the motion quality of our method. Secondly, triangle meshes cannot model dynamic scenes where new objects emerge over time, such as water flowing from a hydrant. Thirdly, generating complex and long-duration motions is challenging and requires significant computational resources and time. Finally, there is a need for a unified evaluation metric to comprehensively assess the attributes of text-to-4D results.

\paragraph{Broader Impacts}
Our method has significant potential to assist in content creation for games, movies, and animations due to the widespread use of mesh in these fields. However, as our method builds upon an existing open-source model and can generate results for any prompt, there is a potential risk of misuse for generating fake content that could harm specific entities. We strongly disapprove of using our method to create harmful or malicious content that could adversely affect society.

{
    \small
    \bibliographystyle{ieeenat_fullname}
    \bibliography{main}

\begin{thebibliography}{40}
\providecommand{\natexlab}[1]{#1}
\providecommand{\url}[1]{\texttt{#1}}
\expandafter\ifx\csname urlstyle\endcsname\relax
  \providecommand{\doi}[1]{doi: #1}\else
  \providecommand{\doi}{doi: \begingroup \urlstyle{rm}\Url}\fi

\bibitem[An et~al.(2023)An, Zhang, Yang, Gupta, Huang, Luo, and Yin]{an2023latent}
Jie An, Songyang Zhang, Harry Yang, Sonal Gupta, Jia-Bin Huang, Jiebo Luo, and Xi Yin.
\newblock Latent-shift: Latent diffusion with temporal shift for efficient text-to-video generation.
\newblock \emph{arXiv preprint arXiv:2304.08477}, 2023.

\bibitem[Bahmani et~al.(2023)Bahmani, Skorokhodov, Rong, Wetzstein, Guibas, Wonka, Tulyakov, Park, Tagliasacchi, and Lindell]{Bahmani20234DfyTG}
Sherwin Bahmani, Ivan Skorokhodov, Victor Rong, Gordon Wetzstein, Leonidas Guibas, Peter Wonka, S. Tulyakov, Jeong~Joon Park, Andrea Tagliasacchi, and David~B. Lindell.
\newblock 4d-fy: Text-to-4d generation using hybrid score distillation sampling.
\newblock \emph{ArXiv}, abs/2311.17984, 2023.

\bibitem[Ben-Chen et~al.(2009)Ben-Chen, Weber, and Gotsman]{BenChen2009SpatialDT}
Mirela Ben-Chen, Ofir Weber, and Craig Gotsman.
\newblock Spatial deformation transfer.
\newblock In \emph{Symposium on Computer Animation}, 2009.

\bibitem[Blattmann et~al.(2023)Blattmann, Rombach, Ling, Dockhorn, Kim, Fidler, and Kreis]{blattmann2023align}
Andreas Blattmann, Robin Rombach, Huan Ling, Tim Dockhorn, Seung~Wook Kim, Sanja Fidler, and Karsten Kreis.
\newblock Align your latents: High-resolution video synthesis with latent diffusion models.
\newblock In \emph{Proceedings of the IEEE/CVF Conference on Computer Vision and Pattern Recognition}, pages 22563--22575, 2023.

\bibitem[Ge et~al.(2023)Ge, Nah, Liu, Poon, Tao, Catanzaro, Jacobs, Huang, Liu, and Balaji]{ge2023preserve}
Songwei Ge, Seungjun Nah, Guilin Liu, Tyler Poon, Andrew Tao, Bryan Catanzaro, David Jacobs, Jia-Bin Huang, Ming-Yu Liu, and Yogesh Balaji.
\newblock Preserve your own correlation: A noise prior for video diffusion models.
\newblock In \emph{Proceedings of the IEEE/CVF International Conference on Computer Vision}, pages 22930--22941, 2023.

\bibitem[Guo et~al.(2023)Guo, Liu, Shao, Laforte, Voleti, Luo, Chen, Zou, Wang, Cao, and Zhang]{threestudio2023}
Yuan-Chen Guo, Ying-Tian Liu, Ruizhi Shao, Christian Laforte, Vikram Voleti, Guan Luo, Chia-Hao Chen, Zi-Xin Zou, Chen Wang, Yan-Pei Cao, and Song-Hai Zhang.
\newblock threestudio: A unified framework for 3d content generation.
\newblock \url{https://github.com/threestudio-project/threestudio}, 2023.

\bibitem[Hartigan and Wong(1979)]{Hartigan1979AKC}
John~A. Hartigan and M.~Anthony. Wong.
\newblock A k-means clustering algorithm.
\newblock 1979.

\bibitem[Ho and Salimans(2022)]{ho2022classifier}
Jonathan Ho and Tim Salimans.
\newblock Classifier-free diffusion guidance.
\newblock \emph{arXiv preprint arXiv:2207.12598}, 2022.

\bibitem[Ho et~al.(2022)Ho, Chan, Saharia, Whang, Gao, Gritsenko, Kingma, Poole, Norouzi, Fleet, et~al.]{ho2022imagen}
Jonathan Ho, William Chan, Chitwan Saharia, Jay Whang, Ruiqi Gao, Alexey Gritsenko, Diederik~P Kingma, Ben Poole, Mohammad Norouzi, David~J Fleet, et~al.
\newblock Imagen video: High definition video generation with diffusion models.
\newblock \emph{arXiv preprint arXiv:2210.02303}, 2022.

\bibitem[Hu et~al.(2021)Hu, Shen, Wallis, Allen-Zhu, Li, Wang, Wang, and Chen]{hu2021lora}
Edward~J Hu, Yelong Shen, Phillip Wallis, Zeyuan Allen-Zhu, Yuanzhi Li, Shean Wang, Lu Wang, and Weizhu Chen.
\newblock Lora: Low-rank adaptation of large language models.
\newblock \emph{arXiv preprint arXiv:2106.09685}, 2021.

\bibitem[Kerbl et~al.(2023)Kerbl, Kopanas, Leimkuehler, and Drettakis]{Kerbl20233DGS}
Bernhard Kerbl, Georgios Kopanas, Thomas Leimkuehler, and George Drettakis.
\newblock 3d gaussian splatting for real-time radiance field rendering.
\newblock \emph{ACM Transactions on Graphics (TOG)}, 42:\penalty0 1 -- 14, 2023.

\bibitem[Laine et~al.(2020)Laine, Hellsten, Karras, Seol, Lehtinen, and Aila]{laine2020modular}
Samuli Laine, Janne Hellsten, Tero Karras, Yeongho Seol, Jaakko Lehtinen, and Timo Aila.
\newblock Modular primitives for high-performance differentiable rendering.
\newblock \emph{ACM Transactions on Graphics (ToG)}, 39\penalty0 (6):\penalty0 1--14, 2020.

\bibitem[Liao et~al.(2022)Liao, Yang, Saito, Pons-Moll, and Zhou]{Liao2022SkeletonfreePT}
Zhouyingcheng Liao, Jimei Yang, Jun Saito, Gerard Pons-Moll, and Yang Zhou.
\newblock Skeleton-free pose transfer for stylized 3d characters.
\newblock In \emph{European Conference on Computer Vision}, 2022.

\bibitem[Ling et~al.(2023)Ling, Kim, Torralba, Fidler, and Kreis]{Ling2023AlignYG}
Huan Ling, Seung~Wook Kim, Antonio Torralba, Sanja Fidler, and Karsten Kreis.
\newblock Align your gaussians: Text-to-4d with dynamic 3d gaussians and composed diffusion models.
\newblock \emph{ArXiv}, abs/2312.13763, 2023.

\bibitem[Liu et~al.(2023)Liu, Shi, Chen, Zhang, Xu, Wei, Chen, Zeng, Gu, and Su]{liu2023one}
Minghua Liu, Ruoxi Shi, Linghao Chen, Zhuoyang Zhang, Chao Xu, Xinyue Wei, Hansheng Chen, Chong Zeng, Jiayuan Gu, and Hao Su.
\newblock One-2-3-45++: Fast single image to 3d objects with consistent multi-view generation and 3d diffusion.
\newblock \emph{arXiv preprint arXiv:2311.07885}, 2023.

\bibitem[Magnenat-Thalmann et~al.(1989)Magnenat-Thalmann, Laperri{\`e}re, and Thalmann]{MagnenatThalmann1989JointdependentLD}
Nadia Magnenat-Thalmann, Richard Laperri{\`e}re, and Daniel Thalmann.
\newblock Joint-dependent local deformations for hand animation and object grasping.
\newblock 1989.

\bibitem[Meyer et~al.(2002)Meyer, Desbrun, Schr{\"o}der, and Barr]{Meyer2002DiscreteDO}
Mark Meyer, Mathieu Desbrun, Peter Schr{\"o}der, and Alan~H. Barr.
\newblock Discrete differential-geometry operators for triangulated 2-manifolds.
\newblock In \emph{International Workshop on Visualization and Mathematics}, 2002.

\bibitem[Mildenhall et~al.(2020)Mildenhall, Srinivasan, Tancik, Barron, Ramamoorthi, and Ng]{Mildenhall2020NeRF}
Ben Mildenhall, Pratul~P. Srinivasan, Matthew Tancik, Jonathan~T. Barron, Ravi Ramamoorthi, and Ren Ng.
\newblock Nerf: Representing scenes as neural radiance fields for view synthesis.
\newblock \emph{Communications of the ACM}, 65:\penalty0 99 -- 106, 2020.

\bibitem[Nichol and Dhariwal(2021)]{nichol2021improved}
Alexander~Quinn Nichol and Prafulla Dhariwal.
\newblock Improved denoising diffusion probabilistic models.
\newblock In \emph{International conference on machine learning}, pages 8162--8171. PMLR, 2021.

\bibitem[Pinkall and Polthier(1993)]{Pinkall1993ComputingDM}
Ulrich Pinkall and Konrad Polthier.
\newblock Computing discrete minimal surfaces and their conjugates.
\newblock \emph{Exp. Math.}, 2:\penalty0 15--36, 1993.

\bibitem[Poole et~al.(2022)Poole, Jain, Barron, and Mildenhall]{poole2022dreamfusion}
Ben Poole, Ajay Jain, Jonathan~T Barron, and Ben Mildenhall.
\newblock Dreamfusion: Text-to-3d using 2d diffusion.
\newblock \emph{arXiv preprint arXiv:2209.14988}, 2022.

\bibitem[Qiu et~al.(2023)Qiu, Chen, Gu, Zuo, Xu, Wu, Yuan, Dong, Bo, and Han]{Qiu2023RichDreamerAG}
Lingteng Qiu, Guanying Chen, Xiaodong Gu, Qi Zuo, Mutian Xu, Yushuang Wu, Weihao Yuan, Zilong Dong, Liefeng Bo, and Xiaoguang Han.
\newblock Richdreamer: A generalizable normal-depth diffusion model for detail richness in text-to-3d.
\newblock \emph{ArXiv}, abs/2311.16918, 2023.

\bibitem[Rombach et~al.(2022)Rombach, Blattmann, Lorenz, Esser, and Ommer]{rombach2022high}
Robin Rombach, Andreas Blattmann, Dominik Lorenz, Patrick Esser, and Bj{\"o}rn Ommer.
\newblock High-resolution image synthesis with latent diffusion models.
\newblock In \emph{Proceedings of the IEEE/CVF conference on computer vision and pattern recognition}, pages 10684--10695, 2022.

\bibitem[Shi et~al.(2024)Shi, Huang, Wang, Bian, Li, Zhang, Zhang, Cheung, See, Qin, Da, and Li]{Shi2024MotionI2VCA}
Xiaoyu Shi, Zhaoyang Huang, Fu-Yun Wang, Weikang Bian, Dasong Li, Y. Zhang, Manyuan Zhang, Ka~Chun Cheung, Simon See, Hongwei Qin, Jifeng Da, and Hongsheng Li.
\newblock Motion-i2v: Consistent and controllable image-to-video generation with explicit motion modeling.
\newblock \emph{ArXiv}, abs/2401.15977, 2024.

\bibitem[Shi et~al.(2023)Shi, Wang, Ye, Long, Li, and Yang]{shi2023mvdream}
Yichun Shi, Peng Wang, Jianglong Ye, Mai Long, Kejie Li, and Xiao Yang.
\newblock Mvdream: Multi-view diffusion for 3d generation.
\newblock \emph{arXiv preprint arXiv:2308.16512}, 2023.

\bibitem[Singer et~al.(2022)Singer, Polyak, Hayes, Yin, An, Zhang, Hu, Yang, Ashual, Gafni, et~al.]{singer2022make}
Uriel Singer, Adam Polyak, Thomas Hayes, Xi Yin, Jie An, Songyang Zhang, Qiyuan Hu, Harry Yang, Oron Ashual, Oran Gafni, et~al.
\newblock Make-a-video: Text-to-video generation without text-video data.
\newblock In \emph{The Eleventh International Conference on Learning Representations}, 2022.

\bibitem[Singer et~al.(2023)Singer, Sheynin, Polyak, Ashual, Makarov, Kokkinos, Goyal, Vedaldi, Parikh, Johnson, and Taigman]{Singer2023TextTo4DDS}
Uriel Singer, Shelly Sheynin, Adam Polyak, Oron Ashual, Iurii Makarov, Filippos Kokkinos, Naman Goyal, Andrea Vedaldi, Devi Parikh, Justin Johnson, and Yaniv Taigman.
\newblock Text-to-4d dynamic scene generation.
\newblock In \emph{International Conference on Machine Learning}, 2023.

\bibitem[Sohl-Dickstein et~al.(2015)Sohl-Dickstein, Weiss, Maheswaranathan, and Ganguli]{SohlDickstein2015DeepUL}
Jascha~Narain Sohl-Dickstein, Eric~A. Weiss, Niru Maheswaranathan, and Surya Ganguli.
\newblock Deep unsupervised learning using nonequilibrium thermodynamics.
\newblock \emph{ArXiv}, abs/1503.03585, 2015.

\bibitem[Song et~al.(2020)Song, Sohl-Dickstein, Kingma, Kumar, Ermon, and Poole]{Song2020ScoreBasedGM}
Yang Song, Jascha~Narain Sohl-Dickstein, Diederik~P. Kingma, Abhishek Kumar, Stefano Ermon, and Ben Poole.
\newblock Score-based generative modeling through stochastic differential equations.
\newblock \emph{ArXiv}, abs/2011.13456, 2020.

\bibitem[Sorkine-Hornung and Alexa(2007)]{SorkineHornung2007AsrigidaspossibleSM}
Olga Sorkine-Hornung and Marc Alexa.
\newblock As-rigid-as-possible surface modeling.
\newblock In \emph{Eurographics Symposium on Geometry Processing}, 2007.

\bibitem[Sterling(2023)]{Zt2vm}
Spencer Sterling.
\newblock Zeroscope text-to-video model.
\newblock [Online], 2023.

\bibitem[Sumner and Popovi{\'c}(2004)]{Sumner2004DeformationTF}
Robert~W. Sumner and Jovan Popovi{\'c}.
\newblock Deformation transfer for triangle meshes.
\newblock \emph{ACM SIGGRAPH 2004 Papers}, 2004.

\bibitem[Wang et~al.(2023{\natexlab{a}})Wang, Yuan, Chen, Zhang, Wang, and Zhang]{Wang2023ModelScopeTT}
Jiuniu Wang, Hangjie Yuan, Dayou Chen, Yingya Zhang, Xiang Wang, and Shiwei Zhang.
\newblock Modelscope text-to-video technical report.
\newblock \emph{ArXiv}, abs/2308.06571, 2023{\natexlab{a}}.

\bibitem[Wang et~al.(2023{\natexlab{b}})Wang, Yang, Tuo, He, Zhu, Fu, and Liu]{Wang2023SwapAI}
Wenjing Wang, Huan Yang, Zixi Tuo, Huiguo He, Junchen Zhu, Jianlong Fu, and Jiaying Liu.
\newblock Swap attention in spatiotemporal diffusions for text-to-video generation.
\newblock 2023{\natexlab{b}}.

\bibitem[Wang et~al.(2023{\natexlab{c}})Wang, Yang, Tuo, He, Zhu, Fu, and Liu]{wang2023videofactory}
Wenjing Wang, Huan Yang, Zixi Tuo, Huiguo He, Junchen Zhu, Jianlong Fu, and Jiaying Liu.
\newblock Videofactory: Swap attention in spatiotemporal diffusions for text-to-video generation.
\newblock \emph{arXiv preprint arXiv:2305.10874}, 2023{\natexlab{c}}.

\bibitem[Wang et~al.(2019)Wang, Aigerman, Kim, Chaudhuri, and Sorkine-Hornung]{Wang2019NeuralCF}
Yifan Wang, Noam Aigerman, Vladimir~G. Kim, Siddhartha Chaudhuri, and Olga Sorkine-Hornung.
\newblock Neural cages for detail-preserving 3d deformations.
\newblock \emph{2020 IEEE/CVF Conference on Computer Vision and Pattern Recognition (CVPR)}, pages 72--80, 2019.

\bibitem[Wang et~al.(2023{\natexlab{d}})Wang, Chen, Ma, Zhou, Huang, Wang, Yang, He, Yu, Yang, et~al.]{wang2023lavie}
Yaohui Wang, Xinyuan Chen, Xin Ma, Shangchen Zhou, Ziqi Huang, Yi Wang, Ceyuan Yang, Yinan He, Jiashuo Yu, Peiqing Yang, et~al.
\newblock Lavie: High-quality video generation with cascaded latent diffusion models.
\newblock \emph{arXiv preprint arXiv:2309.15103}, 2023{\natexlab{d}}.

\bibitem[Wang et~al.(2024)Wang, Lu, Wang, Bao, Li, Su, and Zhu]{wang2024prolificdreamer}
Zhengyi Wang, Cheng Lu, Yikai Wang, Fan Bao, Chongxuan Li, Hang Su, and Jun Zhu.
\newblock Prolificdreamer: High-fidelity and diverse text-to-3d generation with variational score distillation.
\newblock \emph{Advances in Neural Information Processing Systems}, 36, 2024.

\bibitem[Zheng et~al.(2023)Zheng, Li, Nagano, Liu, Hilliges, and Mello]{Zheng2023AUA}
Yufeng Zheng, Xueting Li, Koki Nagano, Sifei Liu, Otmar Hilliges, and Shalini~De Mello.
\newblock A unified approach for text- and image-guided 4d scene generation.
\newblock \emph{ArXiv}, abs/2311.16854, 2023.

\bibitem[Zhou et~al.(2022)Zhou, Wang, Yan, Lv, Zhu, and Feng]{zhou2022magicvideo}
Daquan Zhou, Weimin Wang, Hanshu Yan, Weiwei Lv, Yizhe Zhu, and Jiashi Feng.
\newblock Magicvideo: Efficient video generation with latent diffusion models.
\newblock \emph{arXiv preprint arXiv:2211.11018}, 2022.

\end{thebibliography}
}

\clearpage \appendix 

\begin{figure*}[htbp]
    \centering
    \includegraphics[width=0.65\linewidth]{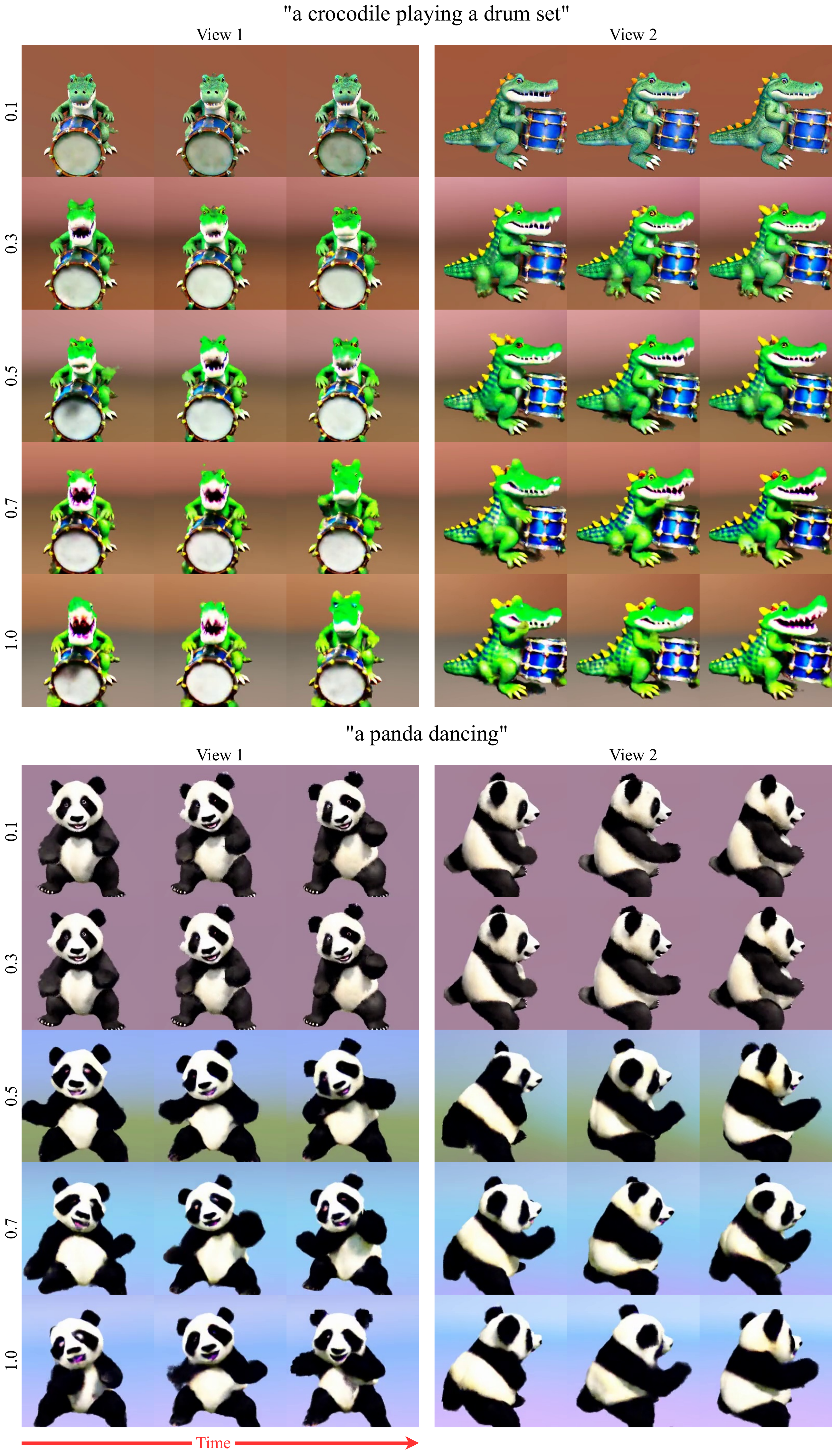}
    \caption{\textbf{Limitations of 4D-fy.} We present the results of varying the video SDS weight from 0.1 to 1.0 in each row. The rendering results are significantly affected by the video diffusion model. Temporally, the geometry at the edges of objects exhibits jitter, and the texture changes, leading to low interframe consistency.}
    \label{fig:08}
\end{figure*}

\section{Discussion of existing methods}
\label{sec:doem}
Since 4D-fy does not decouple the static and dynamic parts, its visual results are influenced by the video diffusion model. To verify that 4D-fy cannot simultaneously achieve pleasing visual results and high motion quality, we designed a set of experiments in which we gradually increased the video SDS loss weight from 0.1 to 1.0 using the same static generating results. The results are shown in Figure \ref{fig:08}. As the weight increases, the appearance of the 4D content degrades rapidly, affecting features such as the face of a crocodile and the abdomen of a panda. In addition to subpar visual results, the geometry is unnaturally deformed, and the texture changes with time, causing significant jitter in the edge regions of objects and leading to low interframe consistency. Furthermore, as mentioned in their paper, 4D-fy requires about 80 GB of VRAM to achieve the reported performance, which restricts the generated video resolution and complicates subsequent editing and the combination of different objects.

\begin{figure*}[htbp]
    \centering
    \includegraphics[width=0.8\linewidth]{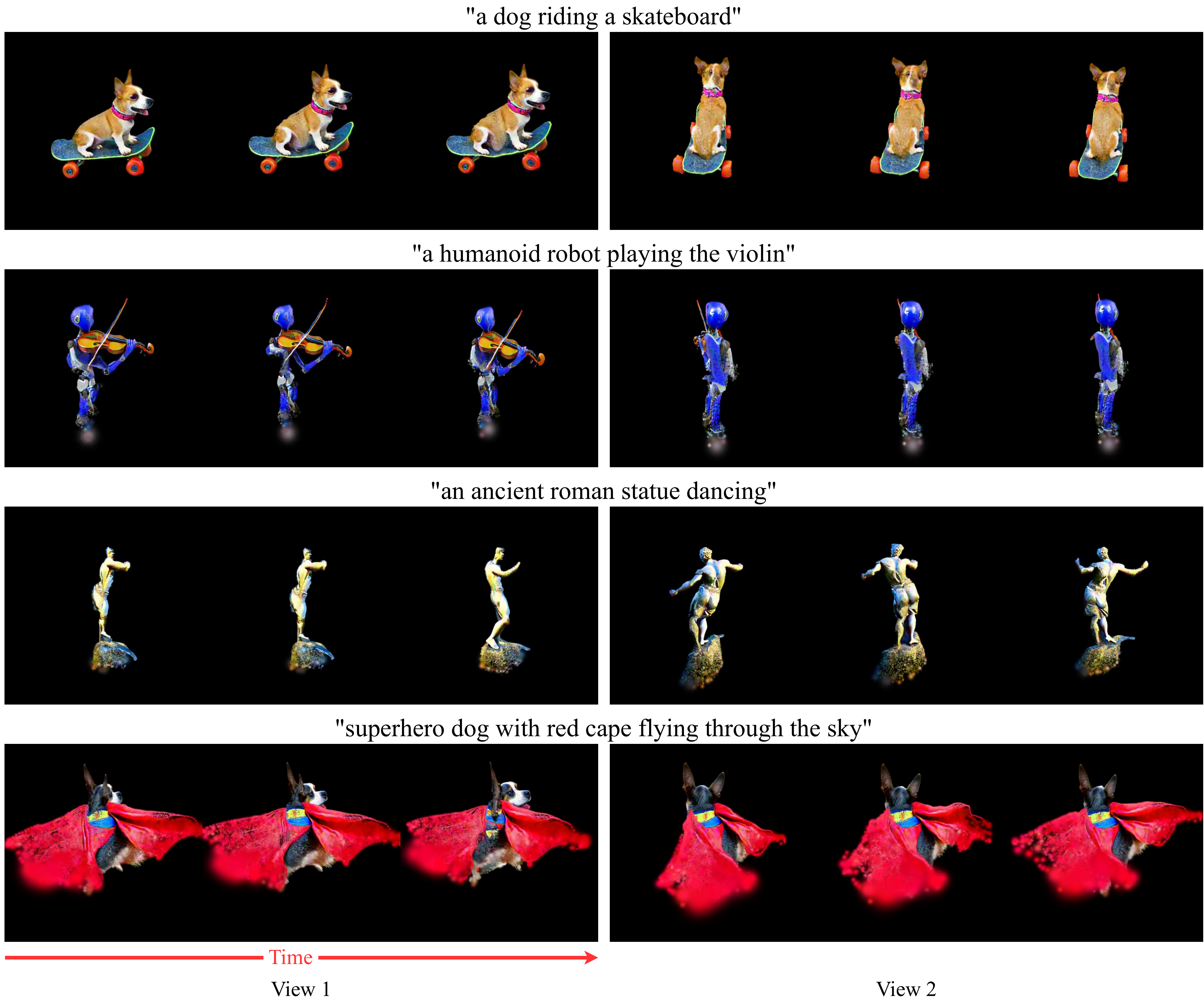}
    \caption{\textbf{Limitations of AYG.} We present some visual results from the official project page of AYG. The rendering results are granular and exhibit poor geometry, resulting in low interframe consistency.}
    \label{fig:09}
\end{figure*}

AYG uses dynamic 3DGS to decouple the static and dynamic parts. However, this often results in granular images due to the inherent particle nature of 3DGS. Additionally, 3DGS tends to overfit specific views, leading to poor geometry. The lack of practical constraints on geometry makes objects unnaturally deform, resulting in low interframe consistency when rendering a video. As shown in Figure \ref{fig:09}, the robot looks distorted at the sencond view and the base of the statue is granulated. Furthermore, AYG does not disentangle geometry and texture, making texture editing difficult.

\section{Preliminary}
\label{sec:pre}

Our \method{} framework incorporates three variants of SDS for text-to-4D generation: single-view SDS, multiview SDS, and video SDS. Additionally, we employ single-view VSD and rigidity regulation in our Generate-Refine-Animate (GRA) algorithm. The following paragraphs explain these components in detail.

\paragraph{SDS (Score Distillation Sampling).}
SDS plays a crucial role in generating 3D objects using diffusion models. In this method, a 3D object represented by learnable parameters $\theta$ is rendered from multiple camera views, and the rendered images $\boldsymbol{x}$ undergo diffusion with a text-to-image diffusion model. The SDS method utilizes the diffusion model's denoiser to construct a gradient, which is then backpropagated through the differentiable rendering process $g$ to update the 3D representation. This update aims to enhance the realism of the rendering results, aligning them with the images generated by the diffusion model. By rendering and incorporating feedback from various camera perspectives, SDS encourages the formation of a geometrically consistent 3D object. The classical SDS gradient \cite{poole2022dreamfusion} is defined as:

\begin{equation}
\begin{aligned}
\nabla_{\boldsymbol{\theta}} & \mathcal{L}_{\mathrm{SV}_\mathrm{SDS}}(\mathbf{x}=g(\boldsymbol{\theta}))=\\
& \mathbb{E}_{t, \boldsymbol{\epsilon}}\left[w(t)\left(\hat{\epsilon}_\phi\left(\mathbf{z}_t, v, t\right)-\boldsymbol{\epsilon}\right) \frac{\partial \mathbf{x}}{\partial \boldsymbol{\theta}}\right].
\end{aligned}
\end{equation}

Here, $t$ represents the diffusion time for perturbing $\boldsymbol{x}$, $w(t)$ is a weighting function, and $\boldsymbol{z}t$ denotes the perturbed rendering. For single-view SDS, $\boldsymbol{x}$ refers to 2D renderings such as images or normal maps. For video SDS, $\boldsymbol{x}$ represents sequences of 2D renderings, such as keyframes. Additionally, $\hat{\boldsymbol{\epsilon}}\phi\left(\mathbf{z}_t, v, t\right)$ represents the diffusion model's denoiser neural network, which predicts the diffusion noise $\epsilon$. It is conditioned on $\boldsymbol{z}_t$, the diffusion time $t$, and a text prompt $v$ for guidance. Classifier-free guidance (CFG) \cite{ho2022classifier} is often employed to enhance the text conditioning. For multiview SDS, camera poses $\boldsymbol{T}$ are included as conditions when calculating the gradient:

\begin{equation}
\begin{aligned}
\nabla_{\boldsymbol{\theta}} & \mathcal{L}_{\mathrm{MV}_\mathrm{SDS}}(\mathbf{x}=g(\boldsymbol{\theta}))=\\
&\mathbb{E}_{t, \boldsymbol{\epsilon},\boldsymbol{T}}\left[w(t)\left(\hat{\epsilon}_\phi\left(\mathbf{z}_t, v, t, \boldsymbol{T}\right)-\boldsymbol{\epsilon}\right) \frac{\partial \mathbf{x}}{\partial \boldsymbol{\theta}}\right].
\end{aligned}
\end{equation}

Here, the expectation is taken over the diffusion time $t$, diffusion noise $\boldsymbol{\epsilon}$, and camera poses $\boldsymbol{T}$. The gradient calculation incorporates the weighted difference between the denoiser's predicted diffusion noise $\hat{\epsilon}_\phi\left(\mathbf{z}_t, v, t, \boldsymbol{T}\right)$ and the actual diffusion noise $\boldsymbol{\epsilon}$. The gradient is then multiplied by the partial derivative of the rendered image $\mathbf{x}$ with respect to the learnable parameters $\boldsymbol{\theta}$. This formulation enables the optimization process to consider the influence of camera poses in achieving a consistent 3D scene representation across multiple views.

\paragraph{VSD (Variational Score Distillation).}
VSD, introduced in \cite{wang2024prolificdreamer}, aims to enhance the appearance of images rendered from the scene. It incorporates a pre-trained text-to-image model \cite{rombach2022high} and employs a fine-tuning scheme to improve image quality beyond what can be achieved by the 3D-aware text-to-image model alone. Following \cite{wang2024prolificdreamer}, we augment the standard SDS gradient with the output of an additional text-to-image diffusion model that undergoes fine-tuning using a low-rank adaptation \cite{hu2021lora} during optimization. The VSD gradient is defined as:

\begin{equation}
\begin{aligned}
\nabla_{\boldsymbol{\theta}} & \mathcal{L}_{\mathrm{SV}_\mathrm{VSD}}(\mathbf{x}=g(\boldsymbol{\theta}))=\\
&\mathbb{E}_{t, \boldsymbol{\epsilon},\boldsymbol{T}}\left[w(t)\left(\hat{\epsilon}_\phi\left(\mathbf{z}_t, v, t\right)-\hat{\epsilon}^{\prime}_\phi\left(\mathbf{z}_t, v, t, \boldsymbol{T}\right)\right) \frac{\partial \mathbf{x}}{\partial \boldsymbol{\theta}}\right].
\end{aligned}
\end{equation}

Here, $\hat{\epsilon}^{\prime}$ represents the noise predicted using a fine-tuned version of the diffusion model that incorporates additional conditioning from the camera extrinsics $\boldsymbol{T}$. The model is fine-tuned using the standard diffusion objective:

\begin{equation}
\min _\theta \mathbb{E}_{t_d, \boldsymbol{\epsilon}, \mathbf{T}}\left[\left\|\boldsymbol{\epsilon}_\phi^{\prime}\left(\mathbf{z}_{t_d} , v, t_d, \mathbf{T}\right)-\boldsymbol{\epsilon}\right\|_2^2\right].
\end{equation}

It's important to note that, similar to 4D-fy, we deviate from the original description of VSD by omitting simultaneous optimization over multiple scene samples.

\paragraph{ARAP (As-Rigid-As-Possible).}
ARAP is a well-known algorithm used for mesh animation and finds applications in various fields. Its fundamental concept involves minimizing an energy function given by:

\begin{equation}
\begin{aligned}
E(\phi, \phi^\prime) =\sum_{i=1}^n \sum_{j \in \mathcal{N}(i)} w_{i j}\left\|\left(\boldsymbol{v}_i^{\prime}-\boldsymbol{v}_j^{\prime}\right)-R_i\left(\boldsymbol{v}_i-\boldsymbol{v}_j\right)\right\|^2.
\end{aligned}
\end{equation}

In this equation, $\phi$ is the static mesh, $\phi^\prime$ is the animatable mesh, $w_{ij}$ represents edge weights, $\boldsymbol{v}$ denotes the vertices of $\phi$, $\boldsymbol{v}^{\prime}$ represents the vertices of $\phi^\prime$, and $R_i$ represents the optimal rotations of the 1-ring neighborhood cells of vertex $\mathbf{p}_i$. The goal is to find vertices $\boldsymbol{v}^{\prime}$ that minimize the distortion while preserving the rigidity of the mesh. The edge weights $w_{ij}$ are calculated with cotangent weight formula \cite{Pinkall1993ComputingDM, Meyer2002DiscreteDO}:

\begin{equation}
w_{i j}=\frac{1}{2}\left(\cot \alpha_{i j}+\cot \beta_{i j}\right),
\end{equation}

where $\alpha_{i j}$, $\beta_{i j}$ are the angles opposite of the mesh edge $(\boldsymbol{v}_i, \boldsymbol{v}_j)$.

\section{User Study Details}
\label{sec:usd}
The user study consists of 3 parts with a total of 35 questions. In each part, a text prompt and generated videos from different methods are shown for each question. The evaluator should read the text prompt, watch all videos, and then choose the best method for each attribute presented with the questions. The attributes include:

\begin{itemize}
    \item \textbf{Appearance Quality (AQ):} Assess the clarity and visual appeal of the scene from any viewpoint, focusing on the appearance of foreground objects while ignoring background inconsistencies.
    \item \textbf{3D Structure Quality (SQ):} Evaluate the level of detail and realism in the shape of the object from various viewpoints, assessing whether the 3D structure of objects makes sense from different angles.
    \item \textbf{Motion Quality (MQ):} Judge the realism of motion, considering both the amount of motion and how well it is depicted in the video.
    \item \textbf{Text Alignment (TA):} Assess the accuracy of each video in representing the content of the text cue, specifically considering if crucial elements of the cue are represented.
    \item \textbf{Interframe Consistency (IC):} Evaluate the consistency of objects across each frame in the video, considering any jitteriness or geometry distortion when the frame switches.
    \item \textbf{Depth Map Quality (DMQ):} Assess the quality of the depth map rendered from the 4D method, focusing on whether the depth map shows near and far relations correctly and smoothly for the object.
    \item \textbf{Normal Map Quality (NMQ):} Assess the quality of the normal map rendered from the 4D method, focusing on whether the surface of the object shown in the normal map is smooth.
    \item \textbf{Extracted Mesh Quality (EMQ):} Assess the quality of the animatable meshes, focusing on the geometry and texture of the animatable meshes.
    \item \textbf{Overall Preference:} Provide a preference rating for the three videos based on personal preference.
\end{itemize}

In the first part, we show the RGB videos generated with AYG, 4D-fy, and our method randomly. The evaluators need to choose the best method for the attributes of AQ, SQ, MQ, TA, IC, and overall preference. This part contains 13 questions.

In the second part, we show the videos of RGB, depth map, normal map, and animatable meshes with and without texture. Since AYG does not provide source code, these videos are generated with 4D-fy and our method. The prompts used in this part are the same as in Part 1. We show the questions randomly. The evaluators need to choose the best method for the attributes of DMQ, NMQ, and EMQ. This part contains 13 questions.

In the third part, we show the videos of RGB, depth map, normal map, and animatable meshes with and without texture. These videos are generated from the ablation study. The evaluators need to choose the best method for the attributes of AQ, SQ, MQ, TA, IC, DMQ, NMQ, EMQ, and overall preference. This part contains 9 questions.

\begin{figure*}[htbp]
    \centering
    \includegraphics[width=0.8\linewidth]{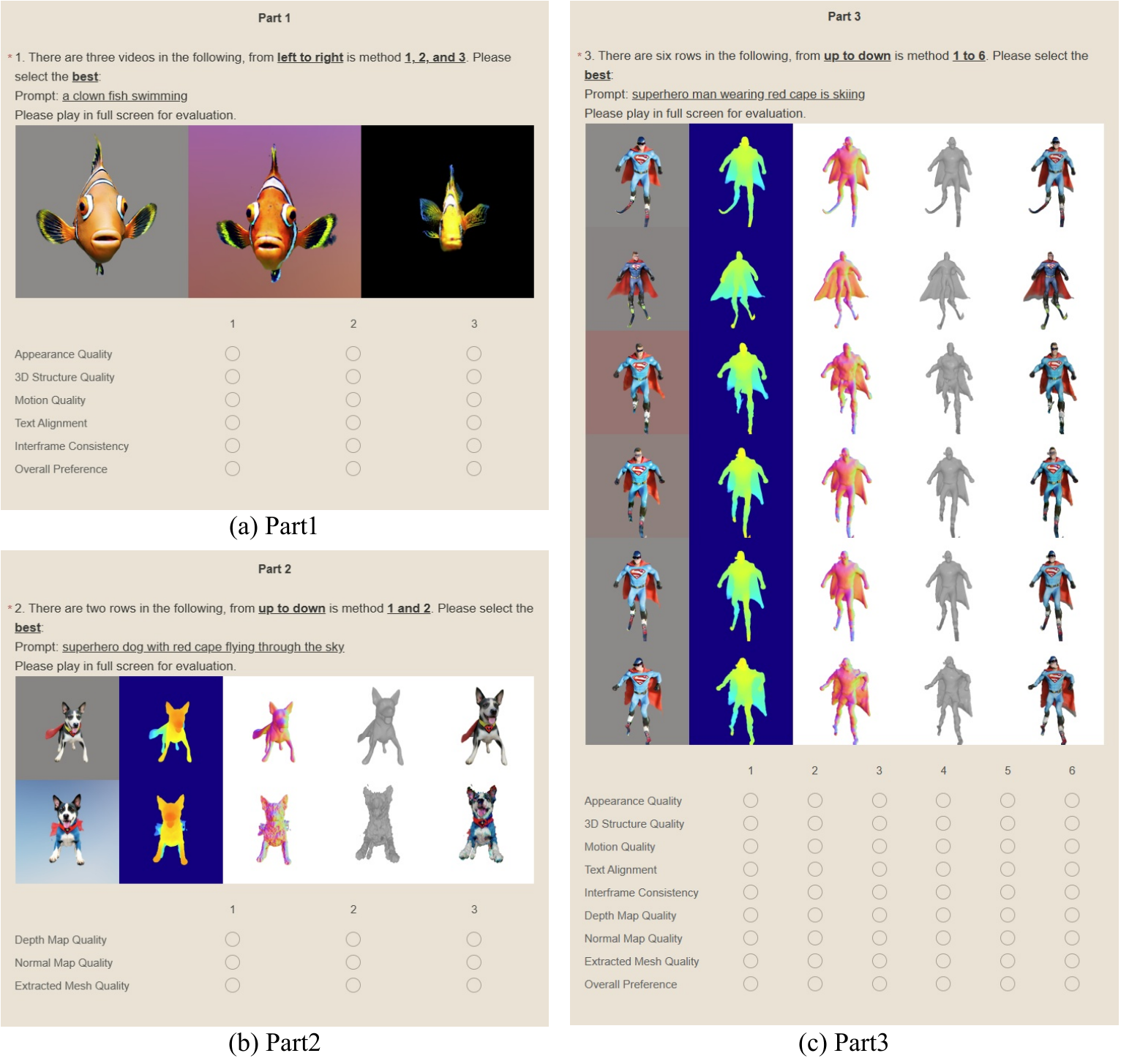}
    \vspace{-2mm}
    \caption{\textbf{User Study Example.} (a) is the example of part 1. (b) is the example of part 2. (c) is the example of part 3.}
    \label{fig:10}
    \vspace{-3mm}
\end{figure*}

The examples of the three parts are shown in Figure \ref{fig:10}.

The prompts used in Part 1 and Part 2 are as follow:
\begin{enumerate}
    \item \textit{"a clown fish swimming"}
    \item \textit{"a dog riding a skateboard"}
    \item \textit{"a donkey running fast"}
    \item \textit{"a humanoid robot playing the violin"}
    \item \textit{"a knight in shining armor holding a sword and shield fighting"}
    \item \textit{"a storm trooper walking forward and vacuuming, best quality, 4k, HD"}
    \item \textit{"an ancient Roman statue dancing, full body, portrait, game, unreal, 4k, HD"}
    \item \textit{"an astronaut riding a horse, best quality, 4k, HD"}
    \item \textit{"an emoji of a baby panda reading a book"}
    \item \textit{"dragon armor fluttering its wings fast"}
    \item \textit{"mage in purple robe dancing, full body, portrait, game, unreal, 4k, HD"}
    \item \textit{"superhero dog with red cape flying through the sky"}
    \item \textit{"tesla trooper standing upright from half squat position, scifi, game, character, photorealistic, 4k, HD"}
\end{enumerate}

The prompts used in Part 3 are as follow:
\begin{enumerate}
    \item \textit{"3D rendering of a fox playing videogame"}
    \item \textit{"a dragon is flying"}
    \item \textit{"an elf playing skateboard"}
    \item \textit{"a cartoon dragon is running"}
    \item \textit{"a horse is walking"}
    \item \textit{"superhero man wearing red cape is skiing"}
    \item \textit{"a dog is running"}
    \item \textit{"a woman is singing"}
    \item \textit{"unicorn running"}
\end{enumerate}

\section{Implementation Details}
\label{sec:id}

\paragraph{Generating.}
When generating the coarse NeRF, the number of optimization iterations is set to 10,000. We adopt the camera sampling technique employed by MVDream \cite{shi2023mvdream}. The sampled camera is oriented towards the origin $(0,0,0)$. For each optimization iteration, we randomly sample the following parameters: field-of-view ($fov$) from a uniform distribution $U(15,60)$, elevation angle ($elv$) from $U(10,45)$, azimuth angle ($azm$) from $U(0,360)$, and camera distance ($cam_{dist}$) from $U(2.5, 3.0)$. The rendering image size for the first 5,000 iterations is $[64, 64]$, and it is changed to $[256, 256]$ for the remaining iterations. The multiview diffusion model proposed in MVDream is used in this stage. The weight for SDS loss is set to 1.0. The percentage of the minimum step for the multiview diffusion model decreases from 0.98 to 0.02 in the first 8,000 iterations and stays at 0.02 for the rest of the iterations. The percentage of the minimum step for the multiview diffusion model decreases from 0.98 to 0.50 in the first 8,000 iterations and stays at 0.50 for the rest of the iterations.

\paragraph{Refining.}
In the geometry refining stage, we first optimize the coarse NeRF. We use the same camera settings as in the generating stage but set the rendering image size to $[128, 128]$ to avoid out-of-memory (OOM) issues since multiple diffusion models are used. Specifically, we use the multiview diffusion model to optimize the rendered RGB images, the normal-depth diffusion model proposed in RichDreamer \cite{Qiu2023RichDreamerAG} to optimize the normal and depth maps, and the single-view diffusion model proposed in \cite{rombach2022high} to optimize the normal map. The weights for all SDS losses are set to 1.0. The percentage of the minimum step for all diffusion models is set to 0.02, and the percentage of the maximum step is set to 0.50.

After 2,000 iterations, we convert the coarse NeRF to DMTet. The camera settings remain the same, but the rendering image size is increased to $[512, 512]$ for high-resolution results. The usage of the diffusion models is the same as in optimizing the coarse NeRF. After another 2,000 iterations, the geometry refining stage is completed.

Since the rigidity regulation requires a specific number of faces in the extracted triangle mesh, we remesh the generated mesh to maintain approximately 20,000 faces to avoid OOM issues. Following this, the texture refinement is performed with the same camera settings. We use the multiview diffusion model with SDS and the single-view diffusion model with VSD to optimize the rendered RGB images. The weights for all SDS losses are set to 1.0. The percentage of the minimum step for all diffusion models remains at 0.02, and the percentage of the maximum step remains at 0.50. The weights for both SDS and VSD losses are set to 1.0. The texture refining part continues for 25,000 iterations.

\paragraph{Animating.}
Before animating the triangle mesh, we convert it to the animatable mesh form mentioned in Section \ref{sec:4r} by clustering it into 80 small regions. The number of keyframes is set to 16. The camera settings are the same as in the generating stage but render a single view each time. We use SDS with the video diffusion model proposed in Zeroscope \cite{Zt2vm, Wang2023ModelScopeTT} to optimize the rendered video. The percentage of the minimum step for video diffusion models is set to 0.02, and the percentage of the maximum step is set to 0.98. The weight for the video SDS loss is set to 0.1. We perform rigidity regulation every 500 iterations of optimization with video SDS. Each time, we run rigidity regulation for up to 500 iterations with $L_{rig}$. This process will end if the change in $L_{rig}$ is smaller than $10^{-7}$.

\section{Limitations}
\label{sec:l}

While our method generates results with high interframe consistency, several limitations should be addressed in future work:

Firstly, the current generation of static objects does not account for subsequent movements, leading to sticking or problematic joints in motion sequences. Incorporating movement considerations in the generation process would enhance coherence and consistency.

Secondly, triangle meshes cannot model dynamic scenes where new objects emerge over time. Since we generate a static mesh first and have no additional geometry generation process in the animating stage, our method cannot introduce new objects to the scene.

Thirdly, current text-to-4D methods struggle with generating complex and long-duration motions, requiring substantial computational resources and time. These limitations stem from challenges in text-to-video methods, such as low image quality, unrealistic motion, short durations, and high costs of training and inference. Advancements in text-to-video models could address these issues and enable more intricate and extended motions.

Lastly, a unified measurement index for evaluating text-to-4D methods is lacking. Current evaluations rely heavily on text-to-3D metrics and user studies. Developing comprehensive evaluation metrics specifically for text-to-4D generation would provide a more accurate and standardized assessment.

Addressing these limitations will advance the field and broaden the applicability of text-to-4D generation techniques.

\section{Additional Qualitative Results}
\label{sec:aqr}
\begin{figure*}[htbp]
    \centering
    \includegraphics[width=0.55\linewidth]{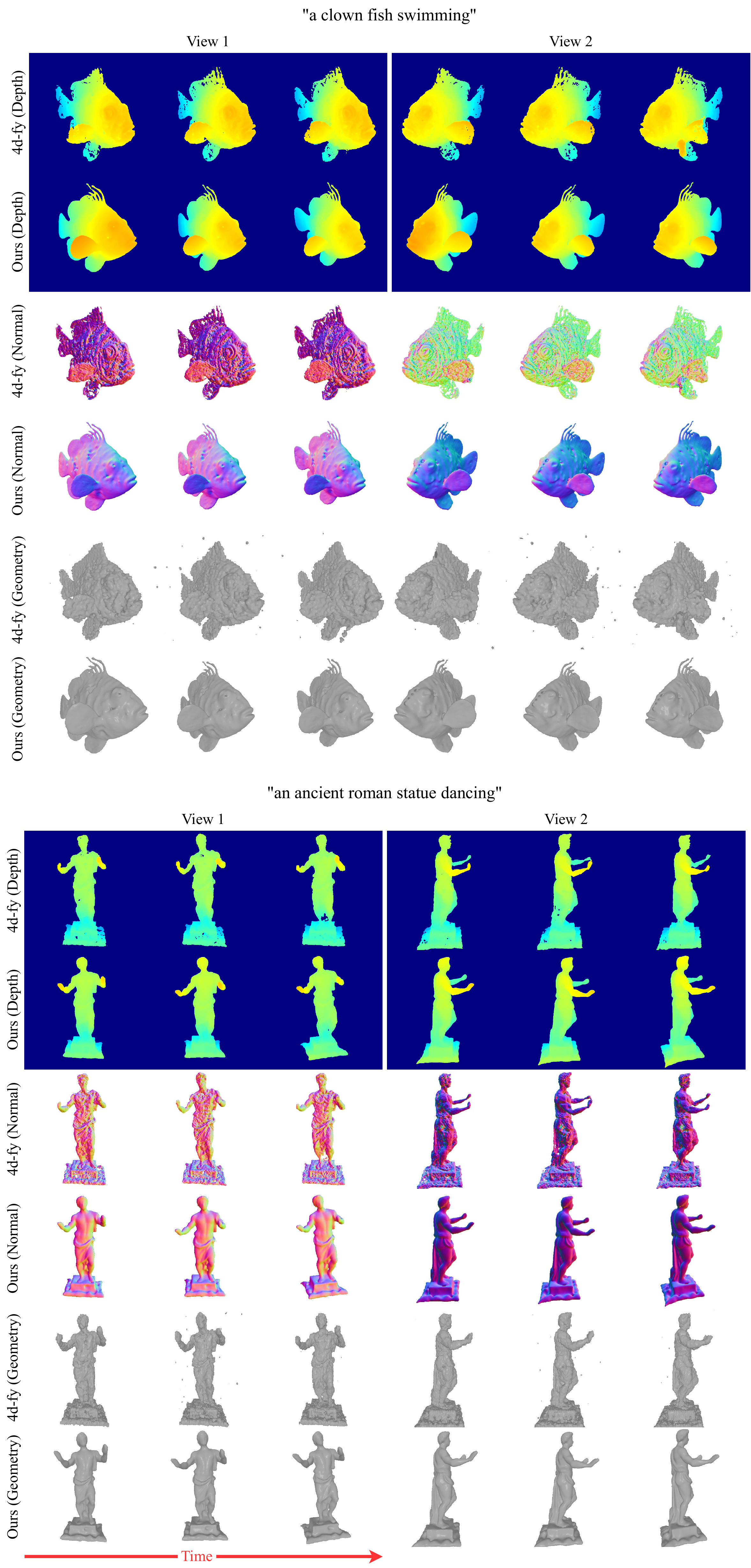}
    \caption{\textbf{Aditional Comparison with 4D-fy.} We compare the depth maps (Depth), normal maps (Normal) and rendered images from mesh without texture (Geometry) between our method with 4D-fy. The results from 4D-fy are noisier compared to those from our method.}
    \label{fig:11}
\end{figure*}

\begin{figure*}[htbp]
    \centering
    \includegraphics[width=0.65\linewidth]{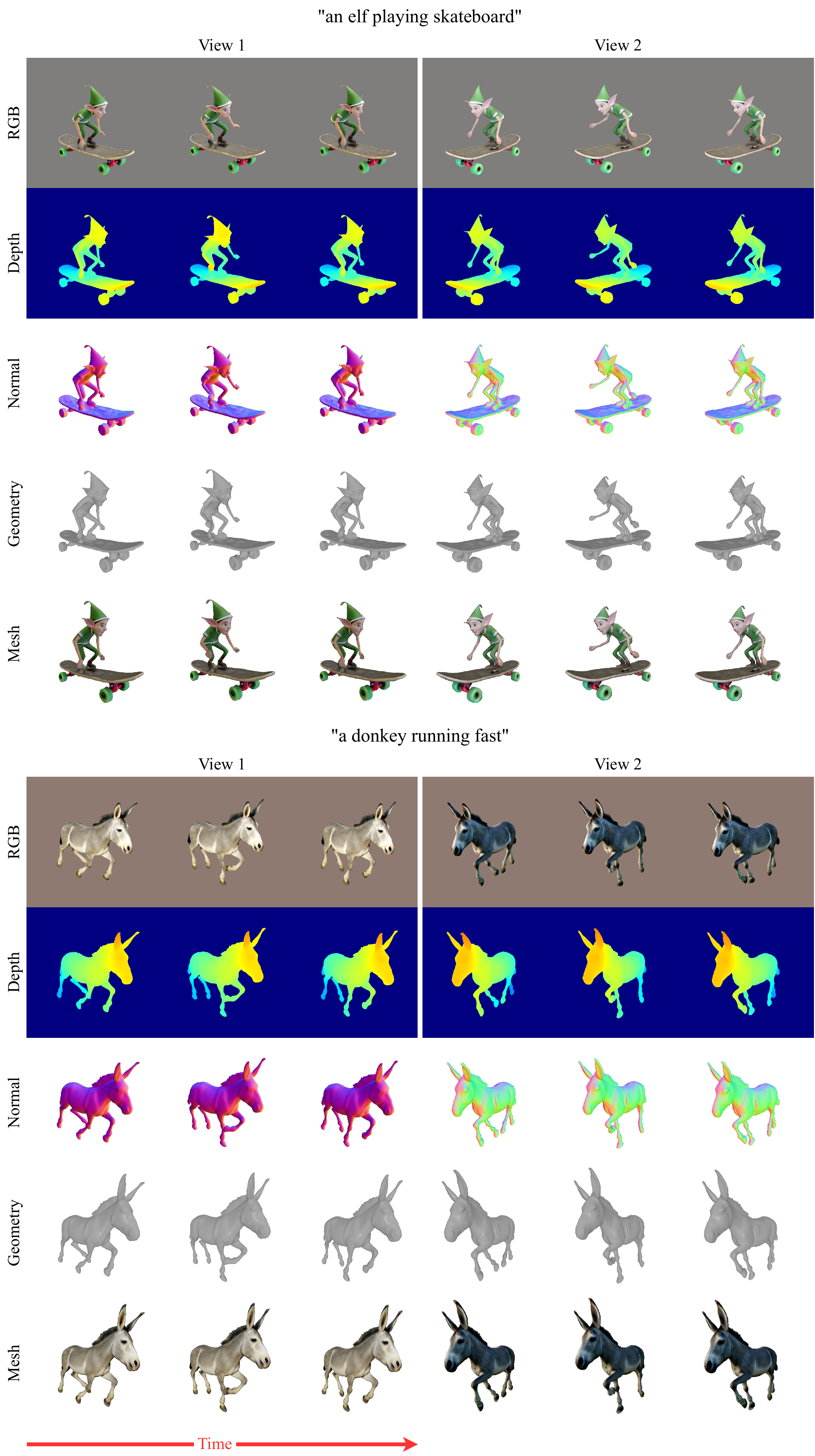}
    \caption{\textbf{text-to-4D Synthesis.} Additional generated results of our method (Part1).}
    \label{fig:12}
\end{figure*}

\begin{figure*}[htbp]
    \centering
    \includegraphics[width=0.65\linewidth]{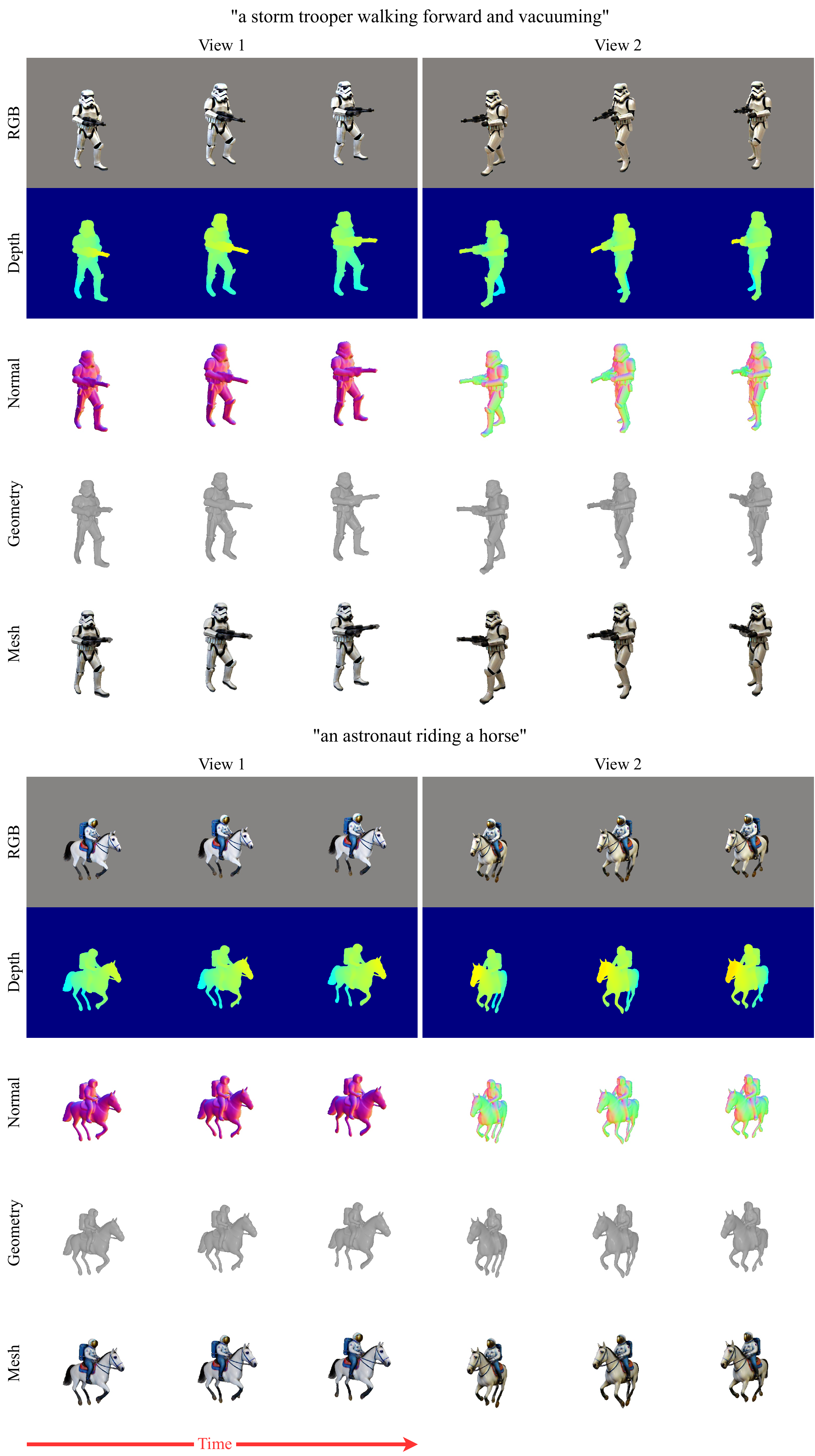}
    \caption{\textbf{text-to-4D Synthesis.} Additional generated results of our method (Part2).}
    \label{fig:13}
\end{figure*}

\begin{figure*}[htbp]
    \centering
    \includegraphics[width=0.75\linewidth]{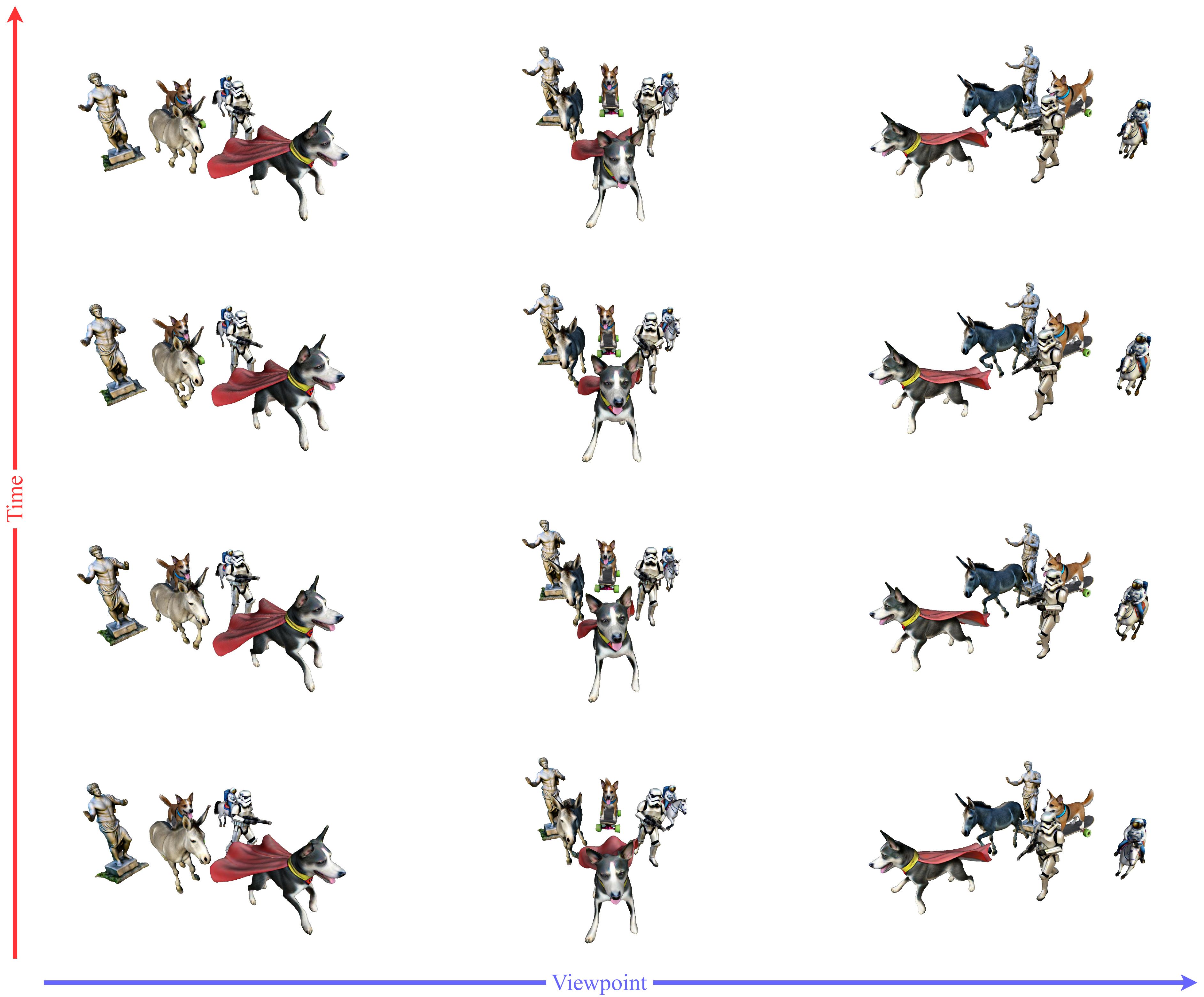}
    \caption{\textbf{Multiple Object Composition.} We show the results of multiple object composition with our method.}
    \label{fig:14}
\end{figure*}

\begin{figure*}[htbp]
    \centering
    \includegraphics[width=0.8\linewidth]{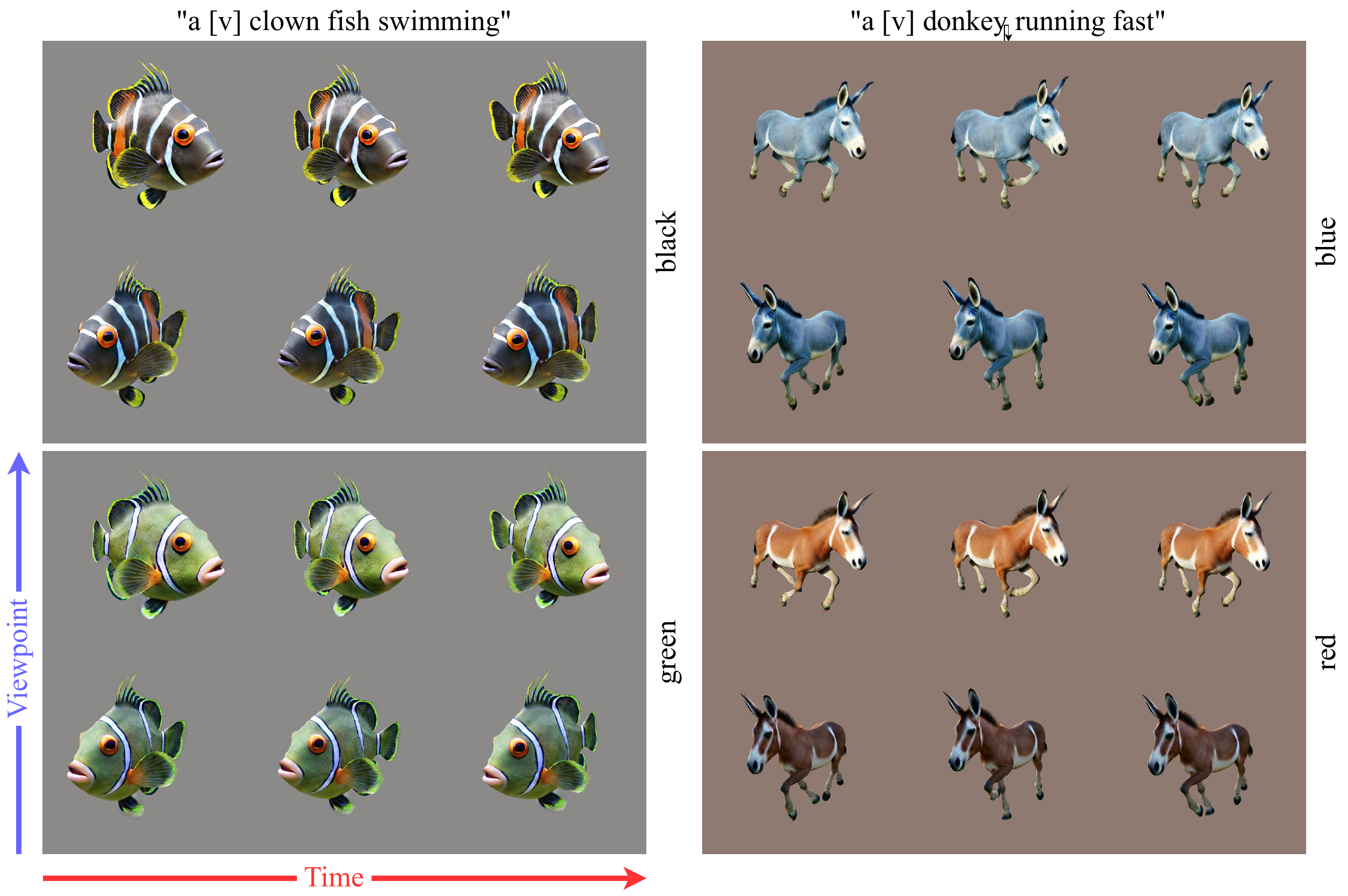}
    \caption{\textbf{Texture Editing} We show the results of texture editing with our method.}
    \label{fig:15}
\end{figure*}

We present an additional comparison with 4D-fy in Figure \ref{fig:11}. Additional generated results of our method are shown in Figures \ref{fig:12} and \ref{fig:13}. Figure \ref{fig:14} demonstrates the results of multiple object compositions. Figure \ref{fig:15} showcases the results of texture editing.

\end{document}